\documentclass[twocolumn,prb,aps,showpacs]{revtex4}
\usepackage{graphics}
\usepackage{dcolumn}
\usepackage{bm}
\usepackage{amsmath}
\usepackage{amssymb}
\newcommand {\Na}{Na$_x$CoO$_2$ }
\newcommand {\Nan}{Na$_x$CoO$_2$}
\newcommand {\NaH}{Na$_x$CoO$_2\cdot y$H$_2$O }
\newcommand {\NaHn}{Na$_x$CoO$_2\cdot y$H$_2$O}
\newcommand {\CN}{$\kappa$-(ET)$_2$Cu$_2$(CN)$_3$ }

\newcommand {\dmit}{$\beta'$-[Pd(dmit)$_2$]$X$ }

\newcommand {\etal}{{\it et al}. }

\usepackage{epsfig}
\usepackage{pictex}
\begin{document}
\title{Ferromagnetism, paramagnetism and a Curie-Weiss metal in an electron doped
Hubbard model on a triangular lattice}
\author{J. Merino$^1$, B. J. Powell$^2$, and Ross H. McKenzie$^2$}
\affiliation{$^1$Departamento de F\'isica Te\'orica de la Materia Condensada,
Universidad Aut\'onoma de Madrid, Madrid 28049, Spain \\
$^2$Department of Physics, University of Queensland, Brisbane, Qld
4072, Australia}
\date{\today}
\begin{abstract}
Motivated by the unconventional properties and rich phase diagram of
\Na we consider the electronic and magnetic properties of a
two-dimensional Hubbard model on an isotropic triangular lattice
doped with electrons away from half-filling. Dynamical mean-field
theory (DMFT) calculations predict that for negative inter-site
hopping amplitudes ($t<0$) and an on-site Coulomb repulsion, $U$,
comparable to the bandwidth, the system displays properties typical
of a weakly correlated metal. In contrast, for $t>0$ a large
enhancement of the effective mass, itinerant ferromagnetism and a
metallic phase with a Curie-Weiss magnetic susceptibility are found
in a broad electron doping range. The different behavior encountered
is a consequence of the larger non-interacting density of states
(DOS) at the Fermi level for $t>0$ than for $t<0$, which effectively
enhances the mass and the scattering amplitude of the
quasiparticles. The shape of the DOS is crucial for the occurrence
of ferromagnetism as for $t>0$ the energy cost of polarizing the
system is much smaller than for $t<0$. Our observation of Nagaoka
ferromagnetism is consistent with the A-type antiferromagnetism
(i.e. ferromagnetic layers stacked antiferromagnetically) observed
in neutron scattering experiments on \Nan. The transport and
magnetic properties measured in Na$_{x}$CoO$_2$ are consistent with
DMFT predictions of a metal close to the Mott insulator and we
discuss the role of Na ordering in driving the system towards the
Mott transition. We propose that ``Curie-Weiss metal'' phase
observed in Na$_x$CoO$_2$ is a consequence of the crossover from
``bad metal'' with incoherent quasiparticles at temperatures $T>T^*$
and Fermi liquid behavior with enhanced parameters below $T^*$,
where $T^*$ is a low energy coherence scale induced by strong local
Coulomb electron correlations. Our analysis also shows that the one
band Hubbard model on a triangular lattice is not enough to describe
the unusual properties of \Na and is used to identify the simplest
relevant model that captures the essential physics in \Nan. We
propose a model which allows for the Na ordering phenomena observed
in the system which, we propose, drives the system close to the Mott
insulating phase even at large dopings.
\end{abstract}
\maketitle

\section{Introduction}

One of the outstanding problems in quantum many-body physics is to
understand quasi-two-dimensional systems in which both
electron-electron interactions and frustration effects are
strong.\cite{Ong&Cava} New exotic phases are expected to result from
the interplay of these effects. Recently a number of materials have
been discovered where the interplay of geometrical frustration and
strong electronic correlations lead to the emergence of
unconventional phases. Examples of such materials include inorganic
materials and organic charge transfer salts such as
$\kappa$-(BEDT-TTF)$_2$Cu$_2$(CN)$_3$ or \dmit (Refs.
\onlinecite{ET-exp} and \onlinecite{dmit-exp} explain this
nomenclature). Antiferromagnets on triangular lattices may be
isotropic as in NiGa$_2$S$_4$,
$\kappa$-(BEDT-TTF)$_2$Cu$_2$(CN)$_3$, and, the material we will
focus on, \Nan, or anisotropic as in Cs$_2$CuCl$_4$,
$\kappa$-(BEDT-TTF)$_2$Cu[N(CN)$_2$]Cl. There is also a great deal
of interest in other frustrated geometries such as the Kagome and
pyrochlore lattices. However, \Na has the, so far, unique property
among strongly correlated triangular lattice compounds, that it can
be doped. In \CN a spin liquid state gives way to superconductivity
when hydrostatic pressure is applied.\cite{CN_spin_liquid} The \dmit
series\cite{Tamura} of compounds, Cs$_2$CuCl$_4$ (Ref.
\onlinecite{Coldea}) and NiGa$_2$S$_4$ (Ref. \onlinecite{Nakatsuji})
may also show spin liquid
behavior. 
Interestingly, the underlying structures of \CN and the \dmit
compounds are very similar to that of NiGa$_2$S$_4$ and \Na with
either the organic molecules, the Ni atoms or the Co atoms arranged
in a triangular geometry. Understanding the physics of these systems
in which both frustration and correlations are present is clearly a
major challenge to theory.

Much attention has been focused on the occurrence of
superconductivity in \NaHn. However, the non-superconducting,
non-hydrated counterpart, \Nan, displays many interesting magnetic
and electronic properties. These include a charge ordered insulating
state in a narrow band of doping around\cite{Foo} $x=0.5$, A-type
antiferromagnetism \cite{Sugiyama} for $x\gtrsim0.75$ and a metallic
phase with a Curie--Weiss susceptibility, $\chi(T) \sim
1/(T+\theta)$, whose magnitude is much larger than the Pauli
susceptibility expected for a weakly interacting metal\cite{Foo}
(the Curie--Weiss metal). This phase also has an extremely large
thermopower (of order $k_B/e$) that is temperature and field
dependent.
\cite{Wang} Foo \etal suggested \cite{Foo} that the metallic state
for $x<0.5$ should be regarded as paramagnetic and the state for
$0.5<x<0.75$ be thought of as a Curie--Weiss metal. These results
should be compared with the measurements of Prabhakaran \etal
\cite{Prabhakaran} which display a different behavior with the
Curie-Weiss behavior strongest at $x=0.5$ (the lowest value of $x$
in their sample set) and the susceptibility becoming less
temperature dependent as $x$ is increased. These differences may be
due to the rather different strength fields used in the two
different experiments,\cite{CWM_foot} a strong dependence on the
direction of the field or to some extrinsic effect. For $x>0.75$
A-type antiferromagnetism (in-plane ferromagnetic ordering stacked
in alternate directions along the $c$-axis) is
observed.\cite{muSR,Bayrakci,Boothroyd,Helme} On the triangular
lattice this raises the interesting possibility of in-plane Nagaoka
ferromagnetism,\cite{Nagaoka} which we will discuss further below.
There is also direct experimental evidence for strong
electron-electron correlations which comes from unconventional
behavior of transport properties. For instance, a Fermi liquid like
resistivity [$\rho(T)=\rho_0+AT^2$] is only observed \cite{Li} for
$T\lesssim1$~K, and the value of the Kadowaki--Woods ratio is
comparable to ruthenates and heavy fermions. \cite{Li,Hussey} The
low temperature scale and the Kadowaki-Woods ratio are strongly
field dependent with the system becoming more weakly correlated with
increasing field.

Interestingly, some of the above properties are reminiscent of the
predictions of the dynamical mean-field theory (DMFT) of the Hubbard
model.\cite{Georges} At half filling ($n=1$) DMFT predicts a Mott
metal-insulator transition for $U\sim W$. In the metallic phase
close to the Mott transition there is a low temperature scale,
$T^*(n)$, at which there is a smooth crossover from a Fermi liquid
to an incoherent ``bad metal". \cite{Georges,Jarrell,Merino}
Transport properties display a crossover from those expected in a
Fermi liquid to properties characteristic of incoherent excitations
as the temperature is increased. Within the ``bad metal'' phase,
that is for $T>T^*$, electrons behave as quasi-localized moments
which leads to a Curie-Weiss behavior of the magnetic
susceptibility. \cite{Georges,Jarrell}  The above discussion
suggests  that DMFT, which treats the local electronic correlations
exactly, may capture the relevant physics needed to describe the
temperature dependence of many of the transport and magnetic
properties of \Nan.

An important first step in studying the properties of \Na is to
consider the two-dimensional triangular lattice Hubbard model. This
is motivated by the arrangement of the Co atoms and because
correlation effects can be singled out and better understood in the
simplest model. Related models such as the $t-J$ model on the
triangular lattice have been recently analyzed using resonating
valence bond (RVB) theory.\cite{Kumar} The implicit assumption in
this kind of approach is that many important aspects of more
realistic models for \Na are already contained in the simpler
triangular lattice model. Our present study, however, leads to the important
conclusion that a single band Hubbard model on a triangular lattice
is not enough to describe the magnetic and electronic properties of \Na
and a more realistic model is proposed.

\begin{figure}
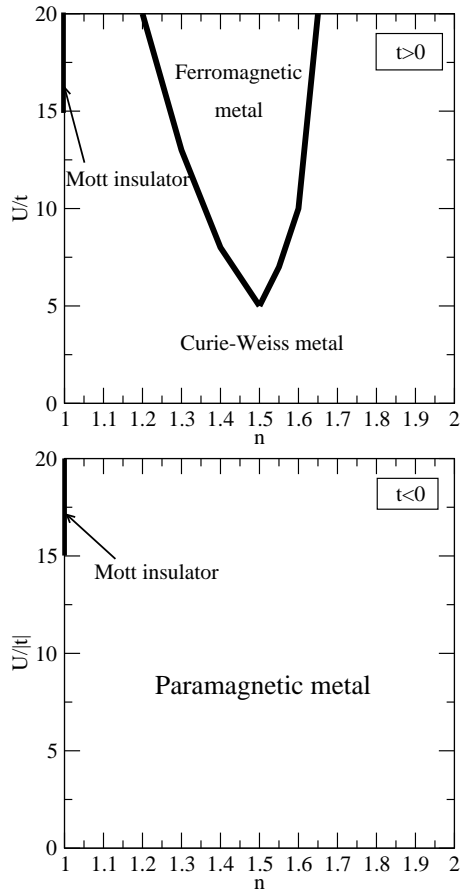

\begin{center}
\epsfig{file=fig1a.eps,width=6.cm,angle=0,clip=}
\epsfig{file=fig1b.eps,width=6.cm,angle=0,clip=}
\end{center}
\caption{Phase diagram for the electron doped Hubbard
model on a triangular lattice
 obtained from DMFT calculations.
$U/|t|$ is the ratio of the on-site Coulomb repulsion
to the absolute value of the hopping matrix element $t$
and $n$ is the average number of electrons per site.
The sign of $t$ has a leads to qualitative
changes in the phase diagram.
 In the $t>0$ case, a ferromagnetic metal and a
metal with Curie-Weiss susceptibility are found. In contrast, for
$t<0$ a paramagnetic metal appears throughout the phase diagram. To
make a direct comparison with \Na one should note that $x=n-1$.}
\label{fig1}
\end{figure}

Our main finding, summarized in Fig. \ref{fig1}, is that DMFT
predicts dramatic changes in the behavior of the Hubbard model on a
triangular lattice when the sign of $t$ is changed. For $t>0$ we
find significant effects due to strong electronic correlations. We
find that in the electron doped triangular lattice a Curie-Weiss
metal and metallic ferromagnetism arise when only local Coulomb
correlations are taken into account. The Curie--Weiss metal
occurs for sufficiently large Coulomb repulsion energies $U \gtrsim
W$ and $t>0$, where $W$ is the bandwidth. In this parameter region,
the local magnetic susceptibility smoothly changes
from the Curie form characteristic of local moments
at high temperature
to Fermi liquid behavior at low temperatures,
$T<T^*(n)\ll\epsilon_F$, where $\epsilon_F$ is the Fermi energy. In
the Curie-Weiss metal the uniform susceptibility is strongly
enhanced by both band narrowing due to correlations
 and the proximity to
the ferromagnetic instability. In contrast, the electron doped
triangular Hubbard model with $t<0$ and $U \gtrsim W$ displays Pauli
paramagnetism weakly enhanced by the Coulomb repulsion; a behavior
typical of weakly correlated metals. Thus our analysis shows how
the non-interacting dispersion has a dramatic effect on the electronic
and magnetic properties of a frustrated Hubbard model. Thus, our work
is relevant to DMFT applications on real materials where similar LDA density
of states appear helping to understand the physics in the more complicated
situations.

The remainder of this paper is organized as follows. In Sec.
\ref{secmodel} we introduce the Hubbard model on a triangular model
discussed above, and then examine the solution of this model on
three sites which already contains some features of, and gives
significant insight too, the solution of the triangular lattice in
the thermodynamic limit. In Sec. \ref{secdmft} we introduce the main
DMFT equations and present arguments which suggest that DMFT, which
is only exact in infinite dimensions, may capture the essential
physics of the Hubbard model on a two-dimensional triangular lattice
due to frustration. In Sec. \ref{secmagnet} we analyze the
temperature dependence of the local and uniform magnetic
susceptibilities for both $t<0$ and $t>0$. In Sec. \ref{secrenor} we
analyze the local correlation effects and the electronic properties
of the model. We report spectral densities and self-energies in both
signs of $t$ and their relation with the different magnetic
susceptibilities found in Sec. \ref{secmagnet}. In Sec.
\ref{secferro} we show that for $t>0$ and electron doping we find
Nagaoka ferromagnetism. In Sec. \ref{secexpt} we discuss the
experimental situation on \Na in the light of our results. In Sec.
\ref{secbeyond} we discuss improvements and extensions to both the
model and the approximation used to study it, we argue that multiple
band models and particularly the effects of ordering of the Na ions
are essential to understand the behaviour observed in \Nan. The
paper ends in Sec. \ref{seccon} in which a summary of main results
is given and our conclusions are drawn.

\section{The Hubbard model on a triangular lattice}
\label{secmodel}

As stated in the introduction, we analyze the magnetic and
electronic properties of a Hubbard model on a triangular lattice,
\begin{eqnarray}
H &=& -t\sum_{\langle ij\rangle,\sigma}\;(c^\dag_{i\sigma} c_{j \sigma} + H.c.) \notag \\
&& + U \sum_{i} n_{i\uparrow} n_{i\downarrow}
-\mu\sum_{i,\sigma}\;n_{i\sigma}, \label{hamilt} \label{tight}
\end{eqnarray}
\noindent where $\mu$, $t$ and $U$ are the chemical potential, the
nearest-neighbor hopping amplitude and the on-site Coulomb repulsion
energies, respectively, $c^{(\dag)}_{i \sigma}$ (creates) destroys
an electron on site $i$ with spin $\sigma$, $\langle \dots\rangle$
indicates that the sum is over nearest neighbours only and the
number operator is $n_{i\sigma}\equiv c^{\dag}_{i \sigma}c_{i
\sigma}$.

 For $U=0$ the
above model gives a tight-binding dispersion
\begin{equation}
\epsilon_{\bf k}=-2t \cos(k_x) -4t \cos\left({\sqrt{3} \over
2} k_y\right) \cos\left({k_x \over 2}\right) - \mu, \label{disper}
\end{equation}
with bandwidth $W=9|t|$.

\subsection{Review of previous work}\label{sec:tri-lat-rev}

For the Hubbard model on the square lattice at half filling, the ground state is
believed to be a Mott insulator with N\'eel order for all values of
$U/t$. The fact that this occurs even for arbitrarily small $U/t$ is
due to the perfect nesting of the non-interacting Fermi surface at
half filling. In contrast, the triangular lattice may exhibit
diverse phases as $U/t$ is varied. Possible ground states that have
been proposed include the Mott insulator, commensurate non-collinear
antiferromagnetism, incommensurate spin-density wave (both metallic
and insulating), spin liquid, superconducting, and metallic states.

In the strong correlation limit, $U\gg t$, there is a gap in the
charge excitation spectrum and the ground state is a Mott insulator.
The spin excitations are described by an antiferromagnetic
 Heisenberg model on the triangular lattice
with exchange constant $J = 4t^2/U$. Numerical calculations from a
range of techniques including exact diagonalisation,\cite{bernu}
variational quantum Monte Carlo,\cite{huse,sorella} coupled cluster
methods,\cite{farnell} and series expansions\cite{singhhuse,weihong}
suggest that this has a magnetically ordered ground state in which
neighbouring spins are rotated by 120 degrees relative to one
another, as in the classical ground state. This agrees with the
predictions of spin wave theory.\cite{chubukov} However, series
expansions calculations suggest that the excitation spectrum is
qualitatively different from that predicted by non-linear spin wave
theory.\cite{zheng2}

At half-filling there is no nesting of the non-interacting Fermi
surface and so for small $U/t$ a metallic state is possible. Exact
diagonalisation calculations of the Drude weight for lattices with
12 sites suggest that there is a first-order transition from a metal
to a Mott insulator when $U \simeq 12.1~t$,\cite{Capone} mean-field
RVB calculations in the thermodynamic limit find a remarkably
similar result with a first order Mott transition occurring at $U
\simeq 12.4~t$.\cite{Powell} When $U \sim 4t$ the effective spin
Hamiltonian in Mott insulating phase  should include ring exchange
terms of order $t^3/U^2$ (Ref. \onlinecite{macdonald}). On the
triangular lattice this can lead to a spin liquid ground
state.\cite{misguich} The possible realisation of a spin liquid in
\CN (Ref. \onlinecite{CN_spin_liquid}) has stimulated the proposal
of specific analytical theories for the ground state in this
regime.\cite{Motrunich,lee+lee} In this parameter regime a
superconducting ground state with $d+id$ symmetry,\cite{flex} or odd
frequency pairing\cite{votja} have also been proposed [it is
therefore interesting to note that odd frequency pairing has also
been discussed on the basis of specific microscopic calculations for
\NaH (Ref. \onlinecite{odd-Na})].

Hartree-Fock\cite{krishnamurthy} and slave boson mean-field
calculations\cite{gazza} have found that near the metal-insulator
phase boundary that incommensurate spin-density wave states are also
stable. Simulated annealing calculations of mean-field solutions
with large possible unit cells found that for $U=2t$ and $U=4t$ that
the most stable states were metals with charge and spin
ordering.\cite{kato2}

For electron doping away from half-filling and $U \gg t$, doubly
occupied states can be projected out and a $t-J$ model can describe
the system. Extensive studies of this model have been
made.\cite{Kumar,baskaran} These studies show that the sign of $t$
has a significant effect on the type of ground state.\cite{Kumar}
RVB calculations\cite{Kumar,baskaran} found that, for $t>0$, $d +
id$ superconductivity was possible for a range of dopings and
suggested a variety of unusual metallic states. Series
expansions\cite{Ogata} have emphasised the competition between
singlet and triplet formation on the triangular lattice in terms of
competition between RVB states and Nagaoka ferromagnetism. This work
concludes that for hole doping the RVB state is favoured by $t>0$
and the Nagaoka state is favoured by $t<0$. Changing the sign of $t$
is equivalent to changing from electron to hole doping, therefore
this result is entirely consistent with the results summarised in
Fig. \ref{fig1}. A more detailed comparison between our results and
those of Koretsune and Ogata\cite{Ogata} will be given in Sec.
\ref{secferro}.

\subsection{Relevance to \Na}

For \Na the LDA bandwidth gives $|t|\sim0.1$~eV.\cite{Singh,LDA} For
electron doping the dispersion (\ref{disper}) leads to a Fermi
surface which is hole-like for $t<0$ and electron-like for $t>0$,
thus $t<0$ is roughly consistent with the band structure
calculations and angle resolved photoemission spectroscopy (ARPES)
experiments for \Nan. However, the band structure of \Na is still
controversial. ARPES \cite{ARPES} seems to suggest that a simple
tight binding model with $t<0$ on the triangular lattice may be
sufficient to describe the $a_{1g}$ bands of \Nan. However, even
within a single band model hopping matrix elements up to at least
3rd nearest neighbors \cite{Rosner,LDA,JohannesEPL} need to be
included in order to correctly reproduce the LDA band structure.
Further, LDA calculations \cite{Singh,LDA} predict the existence of
six small elliptical hole pockets near to the K-points if
$x\lesssim0.6$. It is not clear, at this point, whether or not the
elliptical pockets actually exist.\cite{Lee,Ishida} In contrast to
the $a_{1g}$ bands, the LDA dispersion of the $e_{1g}$ bands is
better fitted to a $t>0$ triangular lattice tight-binding
dispersion\cite{LDA}.

Several attempts \cite{Singh,ARPES} have been made to estimate $U$
for \Nan; all of these find that $U\gg W$. This is much larger than
in the organic charge transfer salts such as \CN or \dmit, in which
$U \gtrsim W$ (Ref. \onlinecite{Mckenzie} and references therein).

It is worth keeping in mind the limitations of this Hamiltonian as a
model for \Nan. For instance, the charge ordered state observed in a
narrow region around $x=0.5$ seems likely to be due to the response
of the electronic system to the ordering of the Na atoms which is
observed even at high temperatures by x-ray diffraction. \cite{Foo}
Therefore one does not expect this phase to appear in purely
electronic models such as those considered here. Nevertheless,
before attempting to understand more complicated models with
realistic band structures (including possibly multiple bands) or
including the effects of Na ordering, it is important to have a firm
understanding of strong correlations in simple frustrated models.
Therefore, despite the remarks above, we present below a DMFT study
of the Hubbard model on a triangular lattice (\ref{tight}). In
section \ref{sec:beyond-Hub} we discuss extensions to the Hubbard
model which allow one to investigate such effects.


\subsection{Kinetic energy frustration}

Geometrically frustrated antiferromagnetism has been studied
extensively and the triangular lattice provides a model system to
investigate. Three of the most widely used quantitative measures of
frustration in antiferromagnets are (i) the degeneracy of the ground
state, (ii) the magnitude of the entropy at low temperatures, and
(iii) the ratio of the ground state energy to the total energy of
maximising the individual interaction energies.\cite{zheng2}

In a non-interacting electron model the only proposal we are aware
of for a quantitative measure of the geometrical frustration of the
kinetic energy is due to Barford and Kim.\cite{barford} They noted
that, for $t>0$, an electron at the bottom of the band does not gain
the full lattice kinetic energy, while a hole at the top of the band
does. They suggested that for tight binding models a measure of the
frustration is then $\Delta = |\epsilon_k^{max}| -
|\epsilon_k^{min}|$, where $\epsilon_k^{max}$ and $\epsilon_k^{min}$
are the energies (relative to the energy of the system with no
electrons) of the top and bottom of the band respectively. This
frustration increases the density of states for positive energies
for $t>0$ (negative energies for $t<0$) which represents an
increased degeneracy and enhances the many-body effects when
the Fermi energy is in this regime. However, perhaps a simpler
measure of the kinetic energy frustration is $W/2z|t|$, where $W$ is
the bandwidth and $z$ is the coordination number of the lattice. The
smaller this ratio, the stronger the frustration is, while for an
unfrustrated lattice $W/2z|t|=1$. For example, on the triangular
lattice the kinetic energy
frustration 
leads to a bandwidth, $W=9 |t|$, instead of
$12 |t|$ as one might na\"ively predict from $W=2z|t|$ since $z=6$.

We show below that geometrical frustration of the kinetic energy
is a key concept required to understand the properties of the
Hubbard model on the triangular lattice. In particular it leads to
particle-hole asymmetry which enhances many-body effects for
electron (hole) doped $t>0$ ($t<0$) lattices.

It should be noted that geometrical frustration of the kinetic
energy is a strictly quantum mechanical effect arising from quantum
interference. This interference arises from hopping around
triangular clusters which will have an amplitude proportional to
$t^3$ which clearly changes sign when $t$ changes sign. In contrast
on the, unfrustrated, square lattice the smallest possible
cluster is the square the amplitude for hopping around a square is
independent of the sign of $t$ as it is proportional to $t^4$.
Barford and Kim\cite{barford} noted that the phase collected by
hopping around a frustrated cluster may be exactly cancelled by the
Aharonov-Bohm phase associated with hopping around the cluster for a
particular choice of applied magnetic field. Thus a magnetic field
may be used to lift the effects of kinetic energy frustration. The
quantum mechanical nature of kinetic energy frustration is in
distinct contrast to geometrical frustration in antiferromagnets
which can occur for purely classical spins.

\subsection{The Hubbard model on a triangular
cluster}\label{a ferromagnet}

\subsubsection{Non-interacting case ($U=0$)}

In order to gain some basic insights into some of the rich physics
associated with the triangular lattice Hubbard model,
we now review the exact solution on a single triangle.
This is the simplest possible model with
kinetic frustration and strongly correlated electrons. This model
already contains some of the features found in the solution of the
Hubbard model on the triangular lattice in the thermodynamic limit
and gives significant insight into that problem. In particular the
solution of this toy problem illustrates why the sign of $t$ is
important and why ferromagnetism arises in the solution to a
problem that only contains kinetic, local and (via superexchange)
antiferromagnetic interactions.

Let us begin by recalling the solution of the two site cluster with
$U=0$ and spinless fermions. Let us label the Wannier functions on
the two sites $|1\rangle$ and $|2\rangle$. For $t>0$ the single
electron ground state is the bonding orbital with wavefunction
$|\Psi\rangle=\frac1{\sqrt2}(|1\rangle+|2\rangle)$ and energy
$E=-t=-|t|$. For $t<0$ the single electron ground state is the
antibonding orbital with
$|\Psi\rangle=\frac1{\sqrt2}(|1\rangle-|2\rangle)$ and $E=t=-|t|$.
(Note that in general changing the sign of $t$ reverses the ordering
of the energy levels.) Thus $W=2t$ for the two site cluster which,
as $z=1$ for the two site lattice, fits with the general expectation
that, for an unfrustrated lattice, the bandwidth is given by $2zt$.

The ground state of the $U=0$ triangular cluster
has some significantly different properties
compared to the two-site cluster.
 For $t>0$ the single electron
ground state is completely bonding with
$|\Psi\rangle=\frac1{\sqrt3}(|1\rangle+|2\rangle+|3\rangle)$ and
$E=-2t=-2|t|$, but for $t<0$ we cannot construct a completely
antibonding solution and so the single electron ground state is
degenerate with
$|\Psi\rangle=\frac1{\sqrt6}(|1\rangle-2|2\rangle+|3\rangle)$ or
$|\Psi\rangle=\frac1{\sqrt6}(|1\rangle+|2\rangle-2|3\rangle)$ and
$E=t=-|t|$. Thus even at the non-interacting level the bandwidth is
reduced ($W=3t$, $z=2$ which implies $W/2zt=3/4$) by the effects of
geometric frustration.

\begin{figure}
\begin{center}
\epsfig{file=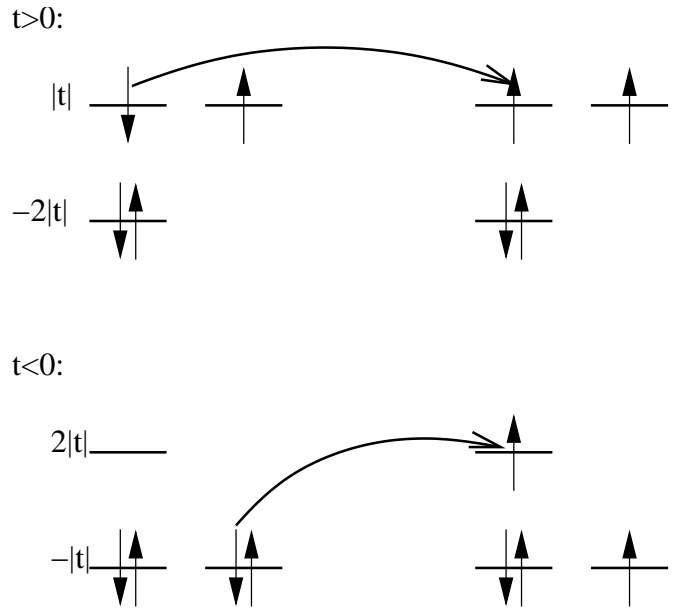,height=8.cm,angle=0,clip=}
\end{center}
\caption{Energy cost in spin polarizing the  three site
triangular cluster with four non-interacting
 electrons. For $t>0$ the ground state is degenerate
and there is no
energy cost in flipping one spin.
In contrast, it costs $3|t|$ to polarize the $t<0$
cluster.
 This indicates that the tendency to ferromagnetism of the
electron doped $t>0$ triangular cluster is much stronger than that
of the cluster with $t<0$ even at the one-electron level.}
\label{fig2}
\end{figure}

Moving to noninteracting spin $1/2$ fermions,  one can already
see how the three site cluster with $t>0$ is
easier to magnetize than the $t<0$ case. In Fig. \ref{fig2} the
energy level structure of the $t<0$ and $t>0$ three site clusters
containing four electrons are displayed. It costs no energy to flip
one spin in the cluster, $\Delta E=E(\uparrow \uparrow \uparrow
\downarrow)-E(\uparrow \downarrow \uparrow \downarrow)=0$, for $t>0$
whereas there is an energy cost of $\Delta E= 3|t|$ associated with
flipping a spin in the $t<0$ case. We will see in section
\ref{secferro} that this tendency to ferromagnetism persists in the
non-interacting triangular lattice.

\subsubsection{Interacting case ($U > 0$)}

\begin{figure}
\begin{center}
\epsfig{file=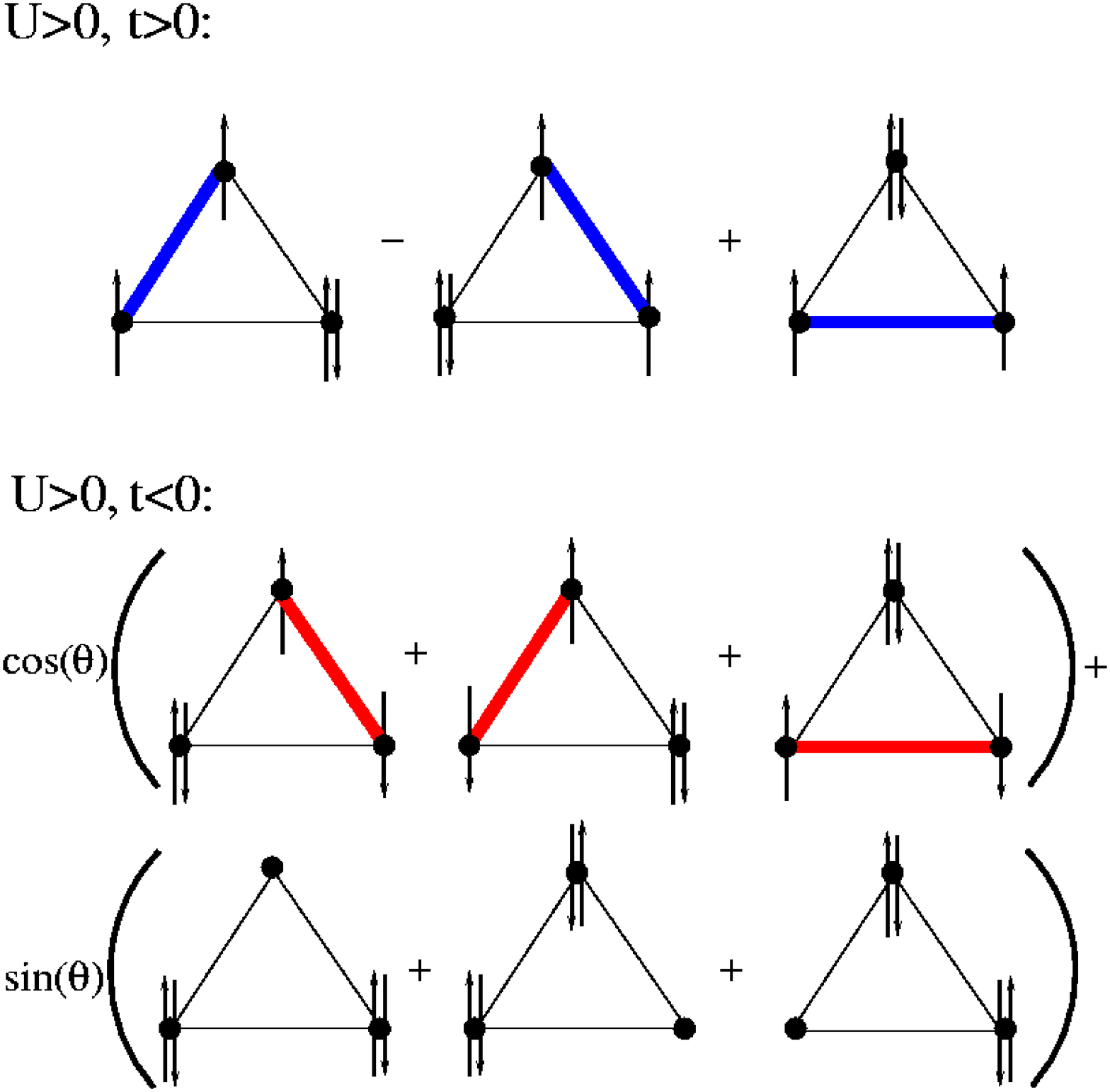,height=8.cm,angle=0,clip=}
\end{center}
\caption{(Color online). A pictorial representation of the ground
states of the Hubbard model on a triangular cluster with four
electrons. Doubly occupied sites are indicated by the presence of
two arrows, one pointing up and the other pointing down, while
unoccupied sites have no arrows. In the cases where we have only one
doubly occupied site the remaining sites with only one electron can
form either a singlet (indicated by a thick red line with arrows on
the two sites pointing in opposite directions) or a triplet
(indicated by a thick blue line with arrows on the two sites
pointing in the same direction). Thus, the middle left triangle
indicates the state
$\frac1{\sqrt2}(|\upharpoonleft\downharpoonright,\uparrow,\downarrow\rangle
- |\upharpoonleft\downharpoonright,\downarrow,\uparrow\rangle)$. The
top left triangle indicates one the three states
$|\downarrow,\downarrow,\upharpoonleft\downharpoonright\rangle$,
$\frac1{\sqrt2}(|\uparrow,\downarrow,\upharpoonleft\downharpoonright\rangle
+ |\downarrow,\uparrow,\upharpoonleft\downharpoonright\rangle)$ or
$|\uparrow,\uparrow,\upharpoonleft\downharpoonright\rangle$.\cite{foot-def}
To keep the figure as simple as possible we have not included
normalization factors in the figure, but the correct normalizations
are given in Eqs. (\ref{psi-}), (\ref{psi+-}), (\ref{psi+0}) and
(\ref{psi++}). The lower part of the figure illustrates
$|\Psi_-\rangle$, the ground state for any non-zero value of $U$ and
$t<0$ [c.f. eqn. (\ref{psi-})] (the ground state is degenerate for
$t<0$ and $U=0$). The upper part of the figure illustrates all three
degenerate states $|\Psi_+^\downarrow\rangle$, $|\Psi_+^0\rangle$
and $|\Psi_+^\uparrow\rangle$ as these only differ by the value of
the $S_z=0$ projection of the triplet. These states are the ground
states for $t<0$ and any value of $U$ [c.f., eqns. (\ref{psi+-}),
(\ref{psi+0}) and  (\ref{psi++})]. $|\Psi_-\rangle$ depends on the
variable $\theta$ which is a function of the ratio $U/t$, given by
eqn. (\ref{thetadef}). $\theta\rightarrow0$ as $U\rightarrow\infty$;
$\theta=\pi/4$ for $U=0$.} \label{fig3}
\end{figure}

For a strongly interacting system the filling factor $n$ will play
an important role in the solution. A non-trivial case in the
triangle relevant to \Na is $n=\frac43$. The $n=\frac53$ case is
trivial as is simply the case of a single (and therefore
non-interacting) hole, while the half filled case will not be
studied in detail in this paper. The relevant basis set is rather
different from that for non-interacting particles and we therefore
illustrate the solution in Fig. \ref{fig6}. To represent this basis
set more precisely we will adopt the notation
$|\alpha,\beta,\gamma\rangle$ where the values of $\alpha$, $\beta$
and $\gamma$ indicate the state on the first, second and third sites
respectively. We allow $\alpha$, $\beta$ and $\gamma$ to take the
values $0$, indicating an unoccupied site, $\uparrow$, indicating a
site singly occupied with a spin up electron, $\downarrow$,
indicating a site singly occupied with a spin down electron, and
$\upharpoonleft\downharpoonright$ which indicates a doubly occupied
site.\cite{foot-def} For $t<0$ we find that for any finite $U$ the
ground state has energy\cite{foot-silly}
\begin{eqnarray}
E_-=\frac12\left(2t+3U-\sqrt{36t^2-4tU+U^2}\right) \label{eqn:E-}
\end{eqnarray}
and is a superposition of singlets
\begin{eqnarray}
|\Psi_-\rangle &=& \frac1{A}\sin\theta \bigg[
|\uparrow\downarrow,\uparrow\downarrow,0\rangle +
|\uparrow\downarrow,0,\uparrow\downarrow\rangle +
|0,\uparrow\downarrow,\uparrow\downarrow\rangle \bigg]\notag\\&& +
\frac1{A}\cos\theta \bigg[ \Big(
|\uparrow\downarrow,\uparrow,\downarrow\rangle -
|\uparrow\downarrow,\downarrow,\uparrow\rangle \Big) \notag\\&&
\hspace*{1.6cm} + \Big(
|\uparrow,\uparrow\downarrow,\downarrow\rangle -
|\downarrow,\uparrow\downarrow,\uparrow\rangle \Big) \label{psi-}
\\&& \hspace*{1.6cm} + \Big( |\uparrow,\downarrow,\uparrow\downarrow\rangle -
|\downarrow,\uparrow,\uparrow\downarrow\rangle \Big) \bigg]\notag,
\end{eqnarray}
where
\begin{eqnarray}
\tan\theta=\frac{U+6t-\sqrt{36t^2-4tU+U^2}}{U-6t+\sqrt{36t^2-4tU+U^2}},\label{thetadef}
\end{eqnarray}
and the normalisation factor is $A=\sqrt{3+3\cos^2\theta}$. Clearly,
$\theta\rightarrow0$ as $U\rightarrow\infty$ and
$\theta\rightarrow-\pi/4$ as $U\rightarrow0$. We sketch this
wavefunction in the lower part of figure \ref{fig3}.

$|\Psi_-\rangle$ is a superposition of the resonating valence bond
state and the states with two doubly occupied sites. The fact that
$\theta\rightarrow0$ as $U\rightarrow\infty$ indicates that the
amplitude for having two doubly occupied sites vanishes in this
limit as one expects. Clearly $|\Psi_-\rangle$ has no net
magnetization, this state leads to the paramagnetic state found in
the thermodynamic limit for $t<0$ (c.f. figure \ref{fig1}).
$|\Psi_-\rangle$ has significant short-range antiferromagnetic spin
fluctuations, which are not captured in the purely local DMFT
treatment that follows, although it is likely that these persist in
the true thermodynamic ground state (see section
\ref{sec:tri-lat-rev}).

For any finite $U$ and $t>0$ the ground state has energy $E_+=-2t+U$
and is threefold degenerate. All three states are spin 1 and three
degenerate states correspond to $S_z=-1,0\textrm{ and }1$. The
respective eigenstates correspond to superpositions of triplets and
are
\begin{eqnarray}
|\Psi_+^\downarrow\rangle&=&
\frac1{\sqrt3}
|\uparrow\downarrow,\downarrow,\downarrow\rangle
-|\downarrow,\uparrow\downarrow,\downarrow\rangle
+|\downarrow,\downarrow,\uparrow\downarrow\rangle,\label{psi+-}\\
|\Psi_+^0\rangle &=&
\frac1{\sqrt6}
\bigg[\Big(
|\uparrow\downarrow,\uparrow,\downarrow\rangle
+|\uparrow\downarrow,\downarrow,\uparrow\rangle \Big) \notag \\&&
\label{psi+0} -\Big(|\uparrow,\uparrow\downarrow,\downarrow\rangle
+|\downarrow,\uparrow\downarrow,\uparrow\rangle \Big)
\\&& \notag +\Big(|\uparrow,\downarrow,\uparrow\downarrow\rangle
+|\downarrow,\uparrow,\uparrow\downarrow\rangle \Big) \bigg]
\end{eqnarray}
and
\begin{eqnarray}
|\Psi_+^\uparrow\rangle &=&
\frac1{\sqrt3}
|\uparrow\downarrow,\uparrow,\uparrow\rangle
-|\uparrow,\uparrow\downarrow,\uparrow\rangle
+|\uparrow,\uparrow,\uparrow\downarrow\rangle.\label{psi++}
\end{eqnarray}
These wavefunctions are illustrated in the upper part of figure
\ref{fig3}.

$|\Psi_+^\downarrow\rangle$, $|\Psi_+^0\rangle$ and
$|\Psi_+^\uparrow\rangle$ display short range ferromagnetic
fluctuations for all finite $U$, in particular the total spin of the
system is a maximum along one axis as the ground state consists of a
superposition of triplets. As the ferromagnetism is associated with
kinetic energy frustration, we have a Nagaoka type ferromagnet for
any finite value of $U$. In the calculations for the infinite
triangular lattice with $t>0$ presented below we will again see this
type of magnetism for some fillings, but with a phase transition
from a Curie-Weiss metal to a Nagaoka ferromagnet at some finite
value of $U$, as is shown in Fig. \ref{fig1}.


In order to quantify the nature and magnitude of the spin
interactions induced by the correlations we evaluate the expectation
value of $\vec{S}_i \cdot \vec{S}_j$ for neigbouring sites on the
triangle. We find that for $t > 0$, this expectation value is 1/12,
whereas for $t < 0$ it is $ -\cos \theta^2 /[3(1 + \cos \theta^2)]$
which becomes more negative with increasing $U$ and tends to -1/6
for $U \gg |t|$. For reference, for the two site Hubbard model, with
two electrons, this expectation value is always negative, regardless
of the sign of $t$, indicating antiferromagnetic spin correlations.
Here, we see that for the triangular cluster, the sign of $t$
determines whether the nearest neighbour spin correlations are
ferromagnetic or antiferromagnetic.

\section{Dynamical Mean-Field Theory}
\label{secdmft}

\subsection{Relevance of DMFT to the two dimensional triangular lattice}

Before presenting our results an important issue to address is the
relevance of DMFT to model (\ref{hamilt}) which describes a
two-dimensional system. DMFT is only exact in infinite dimensions or
for a lattice with infinite coordination number. We will see below
that an important role of the triangular lattice, as compared to
other, non-frustrated, lattices, is not just that the shape of the
bare DOS changes (which is extremely important) but also to make
DMFT a good approximation.

DMFT has become an important tool in the description of strongly
correlated systems. It has provided a realistic description of
transport and dynamical properties of materials such as transition
metal oxides and layered organic superconductors with frustrated
lattices such as $\kappa$-(BEDT-TTF)$_2X$ where $X$ is an anion,
e.g., I$_3$ or Cu$_2$(CN)$_3$. The $\kappa$-(BEDT-TTF)$_2X$ family
have a phase diagram (as a function of pressure, uniaxial stress, or
chemical substitution) in which a superconducting phase is in close
proximity to a Mott insulating phase. In the metallic phase of the
$\kappa$-(BEDT-TTF)$_2X$ there is a temperature scale, $T^*$, at
which there is a smooth crossover from a Fermi liquid to an
incoherent ``bad metal'' [characterized by the absence of a Drude
peak in the frequency dependent conductivity, a resistivity of the
order of the Ioffe-Regel-Mott limit ($\hbar/e^2a$, where $a$ is the
lattice constant), a thermopower of the order of $k_B/|e|$, and a
non-monotonic temperature dependence of the thermopower and
resistivity].\cite{Georges2,Merino} The temperature dependence of
transport properties displays a crossover from coherent Fermi liquid
behavior to incoherent excitations. In the case of the organic
superconductors it has been found that the DMFT of the Hubbard model
gives both a qualitative\cite{Merino} and
quantitative\cite{Georges2} description of this crossover from a
Fermi liquid to a ``bad metal'' in which there are no
quasi-particles.

The crossover temperature scale is related to the destruction of
Kondo screening and Fermi liquid behavior with increasing
temperature (above the Kondo temperature $T_K$) in the Anderson
model (which is the effective model solved in the DMFT equations as
described above). In the Anderson model the conduction electrons are
strongly scattered by a (localised) magnetic impurity for $T>T_K$.
But for $T<T_K$ a singlet forms between the impurity and the
conduction electrons. In the DMFT of the Hubbard model for $T>T^*$
the electrons are quasi-localised and the electrons on the single
site treated exactly strongly scatter those in the bath. However,
for $T<T^*$ transport is coherent and the electrons only scatter one
another weakly, thus Fermi liquid behaviour is regained. This is
why, for example, the temperature dependence of the resistivities of
the Anderson and Hubbard models are so similar.

The success of DMFT in describing the transport properties and the
phase diagram of many organic charge transfer salts down to
temperatures of about 10-20 K (where, for example, superconductivity
becomes important) has been rather puzzling given that these
materials are quasi-two-dimensional and DMFT is only expected to be
a good approximation in the limit of high dimension or co-ordination
number. However, the applicability of DMFT to low-dimensional
systems with large frustration is consistent with the fact that for
frustrated magnetic models a Curie-Weiss law holds down to a much
lower temperature than for unfrustrated
models\cite{Ramirez,Schiffer,Zheng} indicating the presence of well
formed local moments. Recently, Zheng \etal\cite{Zheng} calculated
the temperature dependence of the magnetic susceptibility of the
antiferromagnetic Heisenberg model on an anisotropic triangular
lattice. They found that for models close to the isotropic
triangular lattice (i.e., models with significant magnetic
frustration) that the Curie-Weiss law (which is a mean-field, single
site approximation) holds down to relatively low temperatures.
Deviations from Curie-Weiss behavior result from spatially dependent
correlations. Hence, we expect that a DMFT treatment of the Hubbard
model on the triangular lattice will be a good approximation down to
much lower temperatures than unfrustrated models. Furthermore, in
the ``bad metal'' region magnetic properties such as the uniform
susceptibility and spin relaxation rate, can be described by the
Heisenberg model because the electrons are essentially localized due
to the proximity to the Mott insulating phase. This means that the
susceptibility can be fit to a Curie-Weiss form down to temperatures
much less than the exchange energy $J$. Furthermore, the spin
correlation length of the antiferromagnetic Heisenberg model
increases with temperature $T$ much more slowly for the triangular
lattice than the square lattice.\cite{Elstner} Specifically, at
$T=0.3J$ the spin correlation length is only one lattice constant
for the triangular lattice. In contrast, for the square lattice the
correlation length is about 50 lattice constants, at
$T=0.3J$.\cite{Elstner}

The above arguments have been recently tested by means of cluster
DMFT calculations. These calculations show how for the isotropic
triangular lattice the solution coincides with single site DMFT
(in particular a quasiparticle peak appears at the Fermi energy).
However, as soon as frustration is released a pseudogap opens up
in the one-electron spectra as a result of short range
antiferromagnetic correlations. \cite{Imai}

A further hint that DMFT is a better approximation on the triangular
lattice than on the square lattice comes from the fact that we find
that at half filling our calculations predict that the Mott
transition occurs at $U_c\approx 15 |t|$ (see Figs. \ref{fig1} and
\ref{fig9}). This compares with exact diagonalization studies on 12
site lattices \cite{Capone} which find that the Mott transition
takes place at $U\approx12|t|$. On the square lattice it is known
that perfect nesting means that the ground state is a Mott insulator
for any finite $U$. However, DMFT predicts\cite{Georges} that
$U_c\approx 12|t|$ unless antiferromagnetism is included. Thus
(without including antiferromagnetism) DMFT gives a qualitatively
incorrect result for the (unfrustrated) square lattice, but a
qualitatively correct result for the (frustrated) triangular
lattice.

\subsection{Formalism}

The electronic and magnetic properties of the Hubbard model on the
triangular lattice (\ref{hamilt}) are analyzed below using
DMFT.\cite{Georges} Previously, various properties of doped Mott
insulators on an hypercubic lattice have been explored within DMFT
in the context of the high-$T_c$ superconductors \cite{Jarrell}
using Quantum Monte Carlo (QMC) techniques. Here, we apply DMFT to a
frustrated lattice and use exact diagonalization and Lanczos
techniques to solve the associated Anderson impurity problem
\cite{Caffarel} at finite and zero temperature, respectively.

We will now briefly describe the DMFT formalism focusing on the
relevant equations and quantities of interest in this work. For a
more detailed discussion see, for example, Ref.
\onlinecite{Georges}. DMFT treats the quantum dynamics on a single
site exactly and the remaining lattice sites provide a bath with
which this site interacts. One may map the Hubbard model onto an
Anderson single-impurity model which must be solved
self-consistently. The bath is described through the Weiss field,
$G_{0\sigma}(i\omega_n)$. The iterative procedure starts by solving
the Anderson model for a given choice of $G_{0\sigma}(i\omega_n)$.
From the on-site Greens function, $G_{\sigma}(i \omega_n)$ we may
obtain the self-energy of the system
\begin{equation}
\Sigma_{\sigma}(i \omega_n)=G^{-1}_{0\sigma}(i \omega_n)-G^{-1}_{\sigma}(i \omega_n),
\end{equation}
which is used to describe the lattice propagator
\begin{eqnarray}
G_\sigma(i \omega_n)&=&\sum_{\bf k} G_\sigma({\bf k}, i
\omega_n)\\&=&\sum_{\bf k} {1 \over i \omega_n + \mu -\epsilon_{\bf
k} -\Sigma_\sigma(i \omega_n)},
\end{eqnarray}
where $\mu$ is the chemical potential, $\omega_n=(2n+1)\pi/\beta$ is
a Matsubara fermionic frequency and $\beta=1/k_BT$. The above
procedure is repeated until a self-consistent solution for the
lattice propagator is found.

\subsubsection{Magnetic susceptibilities}

The magnetization of the system under a small magnetic field, $h$,
is computed from
\begin{equation}
m=\mu_B(n_{\uparrow}-n_{\downarrow}),
\end{equation}
where $n_{\uparrow}$ and $n_{\downarrow}$ are obtained from the
DMFT solution at self-consistency.

The uniform (${\bf q}=0$) susceptibility is then obtained
numerically\cite{Laloux} from
\begin{equation}
\chi(T) = \lim_{h \rightarrow 0} {\partial m \over
\partial h }. \label{chi}
\end{equation}

It is interesting to compare the uniform susceptibility
with the local susceptibility obtained from
\begin{equation}
\chi_{loc}(T)= \sum_{\bf q}\chi({\bf q})= {\mu_B^2} \int_0^\beta
\langle S_z(0) S_z(\tau) \rangle d \tau \label{chiloc}
\end{equation}
where $S_z(\tau)=n_\uparrow(\tau)-n_\downarrow(\tau)$.

The local susceptibility in the imaginary frequency axis is given by
\begin{equation}
\chi_{loc}(i\omega_n) = \mu_B^2 \int_0^\beta e^{-i \omega_n \tau}
\langle S_z(0) S_z(\tau) \rangle d\tau.
\label{chilocim}
\end{equation}
The frequency-dependent local magnetic
susceptibility is related\cite{Georges}
to the nuclear magnetic resonance (NMR) Knight
shift $K(T)$ by
\begin{equation}
K(T)  = A \lim_{\omega\rightarrow0} {\textrm{Re}}\chi_{loc}(\omega+i
\eta) \label{knight}
\end{equation}
when the hyperfine interaction $A$ form factor is independent of
wavevector, i.e. in the local limit appropriate to DMFT. The NMR
relaxation rate $1/T_1$ is given by
\begin{equation}
\frac1{T_1 T} \propto \lim_{\omega\rightarrow0} {1 \over \pi}
\frac{\textrm{Im}\chi_{loc}(\omega+i \eta)}{\omega} \label{nmr}
\end{equation}
where $\omega$ is a real frequency and $\eta$ is an arbitrarily small
real number.

\subsubsection{Numerical methods}

Due to the rapid convergence of the solution with the bath size
\cite{Caffarel} it is sufficient to use a discrete set of states
of $n_s=6-10$ sites to model the electronic bath in order to
obtain reliable results. In what follows we use Lanczos
diagonalization on $n_s=8$ sites to calculate zero temperature
properties. In Sec. \ref{secmagnet} we also present results for
high temperatures for which the Lanczos technique is not adequate,
we therefore use exact diagonalization to solve the Anderson
impurity problem. However, as exact diagonalization is more
computationally expensive than Lanczos diagonalization we are
limited to $n_s=6$ for these calculations. Further it is well
known\cite{Georges} that for exact diagonalization there is a low
energy scale $T^{ns}=E_1-E_0$, where $E_0$ and $E_1$ are
respectively the ground state energy and the energy of the first
excited state of the system (recall that there is a finite energy
gap to the first excited state because of the finite bath size),
below which the results of exact diagonalization calculations are
not trustworthy. We therefore present exact diagonalization
results for finite temperatures with $T>T^{ns}$ and Lanczos data
valid in the $T \rightarrow 0$ limit.  Combining the
two methods allows us to have a rather complete description of the
$T$-dependence of various properties.

\section{Temperature Dependence of Magnetic Susceptibility}
\label{secmagnet}

We now analyze the behavior of the uniform and local magnetic
susceptibilities obtained from DMFT for the Hubbard model on the
triangular lattice for both $t>0$ and $t<0$. We find that
the two different signs of the hopping integral lead to very
different magnetic responses (c.f. Fig. \ref{fig1}) and that
Curie-Weiss metallic behavior appears when $t>0$ but not for
$t<0$.

\subsection{Local Magnetic susceptibility}

The degree of localization of the electrons in the triangular
lattice with different signs of $t$ can be explored by computing the
imaginary time, $\tau$, local spin autocorrelation function,
\begin{equation}
 H(\tau) \equiv \langle S_z(0) S_z(\tau)\rangle.
 \label{gtau}
\end{equation}
Such a correlation function can
be related to a spectral density $A(\omega)$ by\cite{silver}
\begin{equation}
 H(\tau) = \frac{1}{2\pi}
\int_{-\infty}^{\infty} d \omega \exp(-\omega \tau)
\frac{A(\omega)}{1 - e^{-\beta \omega}}.
 \label{gtauspec}
\end{equation}
%
If $A(\omega)=-A(-\omega)$ then
\begin{equation}
H(\tau) = \frac{1}{2\pi} \int_0^{\infty} d \omega \frac{A(\omega)
\left[e^{-\omega \tau} + e^{-(\beta - \tau)\omega}\right]}{1 -
e^{-\beta \omega}}. \label{gtauspec2}
\end{equation}
It then follows that $H(\tau)=H(\beta - \tau)$, as one expects for
the correlation function of commuting operators,\cite{rickayzen} and
$H(\tau)$ should be symmetric about $\tau=\beta/2$. The function
$H(\tau)$ is related to the local susceptibility at Matsubara
frequencies by equation (\ref{chilocim}).

The simplest possible form that the frequency dependent local
susceptibility $\chi_{loc}(\omega)$ [compare equations
(\ref{knight}) and (\ref{nmr})] can take is a ``Drude'' type form
with a single relaxation rate, $\Gamma$,
$\chi_{loc}(\omega)=\chi_0/(1 +i \omega/\Gamma)$, leading to a
Lorentzian form for the spectral density. We note that for a single
impurity Anderson model it was found that this Drude form of the
relaxation rate is a good approximation for temperatures larger than
the Kondo temperature.\cite{Hewson}


In a Fermi liquid, $H(\tau) \sim 1/\tau^2$ for $1/\tau \sim 1/\beta
\ll T^*$. In a Mott insulator, $H(\tau)$ decays exponentially to a
non-zero value for long imaginary times $\tau \sim
\beta/2$.\cite{Georges} The ``bad metal'' regime represents
intermediate behaviour, and the value $H(\tau=\beta/2)$ is a measure
of the extent to which the electrons are localised.

\begin{figure}
\begin{center}
\epsfig{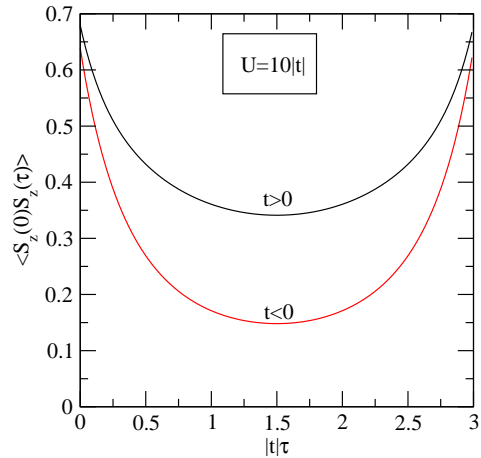}
\end{center}
\caption{(Color online.) The imaginary time local spin
autocorrelation function for both $t>0$ and $t<0$ with $n=1.3$,
$U=10|t|$ and $\beta=3/|t|$. The larger values of $\langle S_z(0)
S_z(\tau=\beta/2) \rangle$ for $t>0$ than for $t<0$ indicates the
greater degree of localization of the electrons for $t>0$. }
\label{fig4}
\end{figure}

In the limit in which $\langle S_z(0) S_z(\tau)\rangle$ becomes
constant one obtains from Eq. (\ref{chiloc})
\begin{equation}
\chi_{loc}(T) = {\mu_B^2 \over T}, \label{approx_chiloc}
\end{equation}
recovering the Curie law for localised magnetic moments. In Fig.
\ref{fig4} we have plotted $\langle S_z(0) S_z(\tau) \rangle$ for
$U=10|t|$, $n=1.3$ at inverse temperature of $\beta=3/|t|$, for
both $t>0$ and $t<0$. Indeed we find that electrons for the $t>0$
triangular lattice (weak $\tau$-dependence) are more localized
than those in the $t<0$ case (strong $\tau$-dependence).
 Both curves are symmetric about $\tau=\beta/2$ as is required.

We find that $\langle S_z(0) S_z(\tau) \rangle$ is not constant for
either sign of $t$ which reflects spin relaxation processes and the
fact that electrons are not completely localized. The actual
behavior at large temperatures is not the pure Curie behavior of Eq.
(\ref{approx_chiloc}) but rather, to a good approximation, can be
fitted to the Curie-Weiss form
\begin{equation}
\chi_{loc}(T) \sim {\mu_{loc}^2 \over T+T^*(n) },
\label{locsu}
\end{equation}
where, $T^*(n)$, is again the coherence scale.

\begin{figure}
\begin{center}
\epsfig{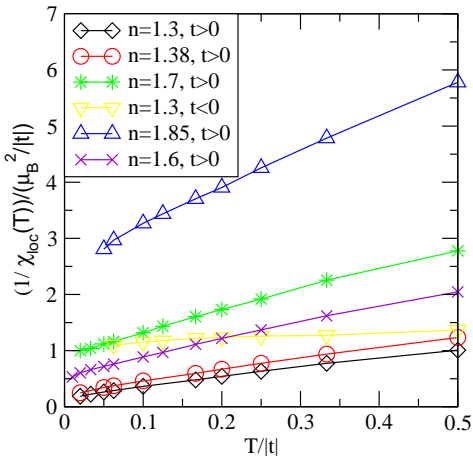}
\end{center}
\caption{(Color online.) The temperature  dependence of the inverse
of the local spin susceptibility for different electron occupations,
$n$. We present results for $U=10|t|$ and for both $t<0$ and $t>0$.
The lines are guides to the eye. For the $t<0$ cases (only $n=1.3$
is shown for clarity) the coherence temperature $T^*\ll|t|$, whereas
for the $t>0$ cases $T^* \sim |t|$. } \label{fig5}
\end{figure}

In Fig. \ref{fig5} we plot $1/\chi_{loc}(T)$ from the actual
numerical results obtained from DMFT for different electron
occupation comparing $t>0$ with the $t<0$ case.
Over a broad temperature range the temperature dependence
is consistent with a Curie-Weiss form. However, the coherence
scale is much larger for
$t<0$ than for $t>0$.


\subsection{Uniform Magnetic susceptibility}

\begin{figure}
\begin{center}
\epsfig{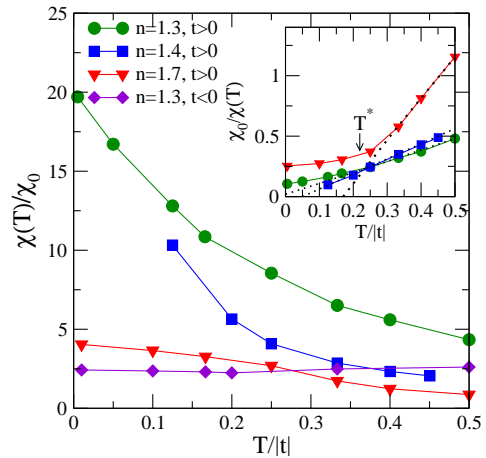}
\end{center}
\caption{(Color online.) Curie-Weiss versus paramagnetic behavior of
the uniform susceptibility, $\chi(T)$, for the Hubbard model on the
triangular lattice with $U=10|t|$. The inset shows $\chi(T)/\chi_0$
(where $\chi_0$ is the zero temperature non-interacting magnetic
susceptibility) comparing the strong $T$-dependence for $t>0$ with
the nearly temperature independent behavior for $t<0$ (Pauli
paramagnetism).  For $n=1.4$ ferromagnetism appears below a critical
temperature, $T_C$.  The main panel displays the inverse of the
uniform susceptibility, $\chi_0/\chi(T)$, showing the Curie-Weiss
$T$-dependence for $t>0$. The arrow indicates the coherence
temperature $T^*$ found from fitting the local susceptibility data
in Fig. \ref{fig5} to Eq. (\ref{locsu}).} \label{fig6}
\end{figure}

In Fig. \ref{fig6} we show the temperature dependence of
$\chi(T)/\chi_0$ for the triangular lattice with $t>0$ and
$U=10|t|$, where $\chi_0$ is the non-interacting uniform
susceptibility of the triangular lattice. We find that the temperature
dependence of the uniform
magnetic susceptibilities are very different for the different
signs of the hopping integral. The magnetic susceptibility
displays Curie-Weiss behavior at large temperatures when $t>0$
whereas for $t<0$, $\chi(T)$ displays the Pauli paramagnetism
characteristic of a weakly correlated metal.

We now discuss the $t>0$ case in more detail.  To be specific, for
$t>0$ and at high temperatures, the uniform susceptibility can be
fitted to the expression
\begin{equation}
\chi(T) \simeq {\mu^2 \over T+\theta(n)}
\label{unisu}
\end{equation}
where $\theta(n)$ and $\mu$ depend on $U$ and $n$.
Note that the temperature scale $\theta(n)$ reduces to
$\theta(n)=-T_C$, where $T_C$ is the critical temperature at which
the ferromagnetic transition occurs for certain values of $n$,
($1.35 \lesssim n \lesssim 1.6$ for $U=10|t|$; see section
\ref{secferro} for details).
The Curie-Weiss behavior described by Eq. (\ref{unisu}) is valid
for large temperatures, $T>T^*(n)$. This is because in this
temperature regime the electronic system behaves like a set of
quasi-localized magnetic moments. For $T<T^*(n)$ the Curie-Weiss
behavior crosses over to a behavior more reminiscent of a
renormalized Fermi liquid with a weak temperature  dependence.
Fitting our results to Eq. (\ref{unisu}) for $T>T^*(n)$, gives
$\theta(n=1.7) \approx -0.17|t|$ and $\theta(n=1.3) \approx 0.01
|t|$ for $n=1.3$. For the ferromagnetic metal appearing for, say,
$n=1.38$, the transition to a ferromagnetic metal occurs at
$T_C=-\theta(n=1.4) \approx 0.05 |t|$. The effective magnetic
moment, $\mu$ obtained from the fitting varies between $\mu=0.7
\mu_B$ for $n=1.3$ and $\mu=0.4 \mu_B$ for $n=1.7$.

\begin{figure}
\begin{center}
\epsfig{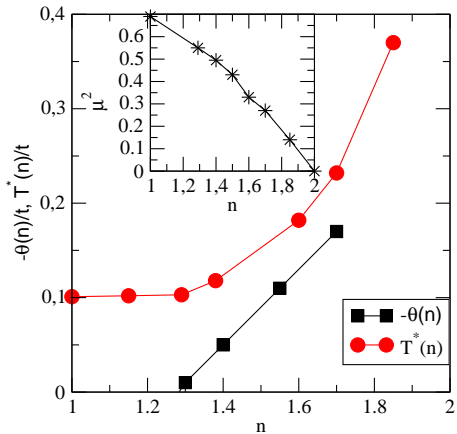}
\end{center}
\caption{(Color online.) Electron occupation dependence of the
coherence scale, $T^*(n)$, the Curie-Weiss scale $\theta(n)$ and the
effective moment in the uniform magnetic susceptibility, $\mu$ for
fixed $U=10|t|$ and $t>0$. These parameters are extracted by fitting
the results of our DMFT calculations of magnetic susceptibilities to
Eqs. (\ref{locsu}) and (\ref{unisu}). For these values of $U$ and
$t$ ferromagnetism is observed in the range $1.3<n<1.6$ (c.f. Fig.
\ref{fig1}). $\theta(n)>0$ for $n<1.3$ and no ferromagnetism is
found at these dopings.  $\mu$ decreases to zero as $n\rightarrow 2$
as the density of unpaired electrons decreases. This appears to be
responsible for the absence of ferromagnetism for $n>1.6$.}
\label{fig7}
\end{figure}

As $n$ is increased one expects that the average moment will be
roughly proportional to the density of unpaired electrons, $2-n$, at
each lattice site. Thus $\mu^2$ is suppressed with increasing $n$
and fixed $U$.  Hence one expects that $\mu^2$ varies between
$\mu=\mu_B$ for $n=1$ (half-filled system) and $\mu=0$ for $n=2$
(filled band). This is consistent with the values of $\mu$ obtained
from the fit of Eq. (\ref{unisu}) to our numerical results (see the
inset of Fig. \ref{fig7}). As the temperature is decreased the
susceptibility changes its behavior so that $\chi_0/\chi(T)$ becomes
less dependent on temperature (this is most obvious for $n=1.7$). A
crossover from Curie-Weiss behavior at high temperatures to Fermi
liquid behavior at low temperatures is encountered. The properties
of the Fermi liquid state (e.g., effective mass renormalization,
susceptibility enhancement) at low temperature will be discussed in
the following section.

In Fig. \ref{fig7} we show a plot of both temperature scales
obtained from fitting our data to the Curie-Weiss laws of Eqs.
(\ref{locsu}) and (\ref{unisu}) to obtain, $T^*(n)$ and $\theta(n)$,
respectively. $T^*(n)$ increases with the electron occupancy. This
is to be expected as the system becomes more weakly correlated in
the $n \rightarrow 2$ limit, remaining coherent at temperatures
comparable with the non-interacting Fermi temperature. On the other
hand, $\theta(n)$ can change sign, which is related to the effective  short range
magnetic coupling, $J(n)$,  present in the uniform susceptibility,
$\chi(T)$, that does not appear in $\chi_{loc}(T)$.  This can be
better understood by splitting $\theta(n)$ into \cite{Georges}
\begin{equation}
\theta(n) \sim T^*(n)+J(n). \label{theta}
\end{equation}
As the system is driven closer to the Mott insulating state: $n
\rightarrow 1$, $\theta(n) \rightarrow J(n)$ as $T^*(n) \rightarrow
0$ which indicates the metal-insulator transition. In this limit the
system behaves as a Heisenberg antiferromagnet with
antiferromagnetic exchange coupling given by $J=4 t^2/U$, with
$U>|t|$. A change of sign in $\theta(n)$ happens at about $n=1.3$ in
Fig. \ref{fig7} signalling a ferromagnetic exchange interaction.
Indeed, this is approximately the doping at which the ferromagnetic
state is found for $U=10|t|$ (c.f. Fig. \ref{fig1}). Thus, as the
electronic occupation increases, $\theta(n)$ changes from positive
(antiferromagnetic) to negative (ferromagnetic) when $J<-T^*$. This
is the threshold ferromagnetic interaction for the ``bad metal'' to
become ferromagnetic. At larger doping values the absolute yields of
the magnetic moment, $\mu$, become suppressed as the number of
unpaired electrons is reduced becoming $\mu \rightarrow 0$ as $n
\rightarrow 2$ which coincides with the destruction of
ferromagnetism.

In contrast to the rather unconventional metallic state found for
$t>0$, the magnetic susceptibility for $t<0$ is Pauli-like,
moderately enhanced by many-body effects. This can be observed in
the inset of Fig. \ref{fig6} where a weak temperature dependence for
all $n$ (for clarity only $n=1.3$ is shown) is found. To be
specific, for $T\ll\epsilon_F$ the low temperature behavior of the
susceptibility for a non-interacting metal is given by
\cite{Ashcroft}
\begin{equation}
{\chi(T) \over \chi_0 } \approx 1+ \left({\rho''(\epsilon_F) \over
\rho(\epsilon_F)}- {\rho'(\epsilon_F)^2  \over
\rho(\epsilon_F)^2}\right){\pi^2 \over 6} k_B^2T^2+ O(T^4),
\label{weak}
\end{equation}
where $\rho(\epsilon)$ is the density of states per spin and the
primes indicate derivatives of the DOS with respect to $\epsilon$.
The non-interacting susceptibility at zero temperature is given by
$\chi_0=2\mu_B^2 \rho(\epsilon_F)$. For $t<0$ a direct evaluation
of the term proportional to $T^2$ gives nearly zero for electron
doping the triangular lattice. Our numerical results agree with
temperature dependence given by Eq. (\ref{weak}), with a moderate
enhancement of $\chi(T)/\chi_0$ suggesting moderate many-body
effects.  Hence, we conclude that the $t<0$ electron doped
triangular lattice behaves as a renormalized paramagnetic metal.

\section{Renormalization of quasiparticles for different signs of $\mathbf t$}
\label{secrenor}

DMFT typically predicts a metal-insulator transition as $U$ is
increased to $U/W \sim 1$ at half-filling.\cite{Georges} As soon as
the system is doped away from half-filling, the system becomes
metallic\cite{Jarrell} with electrons having their mass renormalized
by the Coulomb interaction. We find that the two different densities
of states corresponding to the two different signs of $t$ lead to
different renormalizations of the quasiparticles although $W=9|t|$
is the same for both DOS. This result is in contrast to the
half-filled and unfrustrated cases where the results do not depend
on the sign of $t$. This is because for unfrustrated lattices
particle-hole symmetry is respected, whereas particle-hole symmetry
is broken on the doped triangular lattice. Half-filled frustrated
lattices are {\it not} particle-hole symmetric, but this asymmetry
is suppressed at large $U$, and particularly in the Mott insulating
state.

\begin{figure}
\begin{center}
\epsfig{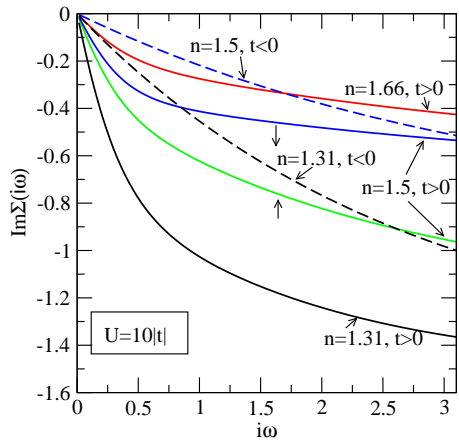}
\end{center}
\caption{(Color online.) Comparison of the imaginary part of the
self-energy   for $t>0$ with the same quantities with $t<0$. Results
are reported for $U=10|t|$ and various electron dopings. The linear
dependence of the self-energy at low energies indicates Fermi liquid
behavior. The slope near $i \omega = 0$ is used to extract the
quasiparticle weight $Z$, shown in Fig. \ref{fig9}.} \label{fig8}
\end{figure}

In order to understand the electronic properties and different
magnetic behavior obtained for $t<0$ and $t>0$ (above),  the
imaginary part of the self-energy along the imaginary axis,
$\text{Im} \Sigma(i\omega)$ for $t<0$ and $t>0$ is computed and
shown in  Fig. \ref{fig8} for $U=10|t|$. The quasiparticle weight
can be extracted from the slope of the self energy,
\begin{eqnarray}
Z=\lim_{\omega\rightarrow0}\left[1-\frac{\partial \text{Im}
\Sigma(i\omega)}{\partial (i\omega)}\right]^{-1}.
\end{eqnarray}
The increasing slope at low frequencies of $\text{Im}
\Sigma(i\omega)$ as $n \rightarrow 1$ indicates a stronger
renormalization of the quasiparticles as the system gets closer to
the Mott metal-insulator transition. This behavior is  apparent for
both signs of $t$. However, for $t>0$ electrons are more strongly
renormalized than for $t<0$. For example, for $n=1.3$ the slope of
$\text{Im} \Sigma(i\omega)$ as $\omega\rightarrow0$ is steeper by a
factor of about 2 for $t>0$ than it is for $t<0$.

\subsection{Effective mass}

\begin{figure}
\begin{center}
\epsfig{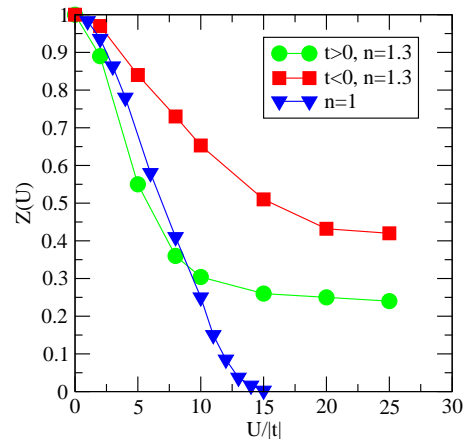}
\end{center}
\caption{(Color online.) Dependence of the quasi-particle weight
 $Z$ on the Coulomb repulsion $U/|t|$ at
various electron dopings. The quasiparticles become more
renormalized as the doping is decreased due to the proximity to the
Mott transition. For a given value of $U$, quasiparticles on the
triangular lattice with $t<0$ are less renormalized than for $t>0$.
In contrast, for the half-filled case ($n=1$), we find the same
behavior of $Z$ for both $t>0$ and $t<0$ with a Mott metal-insulator
transition taking place at a common value $U/W \sim 1.66$.}
 \label{fig9}
\end{figure}

The quasiparticle weight extracted from the self energy as a
function of the Coulomb interaction $U$ is shown in Fig.
\ref{fig9}. $Z$ is compared for $t>0$ and $t<0$ when $n=1.3$. The
effective mass is defined by $m^*=m_b/Z$; where $m_b$ is the
non-interacting band mass. For $U=10|t|$ we find for $t<0$, $m^*=1.5
m_b$ whereas for $t>0$, $m^*=3.4 m_b$. For $n=1$ we find $m^*=4.0
m_b$ independent of the sign of $t$. This is an important point as
it shows that for the same $U/W$ the two different signs of $t$ lead
to different renormalizations of the electrons. Within DMFT this
different renormalization of the quasiparticles can be understood
from the fact that for electron doping the DOS close to the Fermi
energy is larger for $t>0$ than for $t<0$ (c.f., the bare DOS shown
in Fig. \ref{fig6}).

The different renormalization of the electrons found for the
different signs of $t$ in the electron doped system is in contrast
to the very similar renormalization found in the half-filled case
($n=1$) for which $Z$ decreases rapidly and in the same way for both
$t>0$ and $t<0$, leading to a very similar critical value of $U_c
\approx 1.65 W$ at which the Mott transition occurs. This value is
similar to DMFT estimates obtained for the Bethe
lattice\cite{Georges}: $U_c \approx (1.5-1.7)W$.
   The value obtained $U_c=15|t|$ should be compared
with the $U_c\approx12|t|$ obtained from exact diagonalization
calculations for 12 site clusters.\cite{Capone} The similar critical
values appearing at half-filling can be explained from the fact that
as $U \rightarrow W$, the spectral densities at each site are
strongly modified by the Coulomb interaction washing out the fine
details of the bare DOS. In contrast, for the doped case, kinetic
energy effects are important.

\begin{figure}
\begin{center}
\epsfig{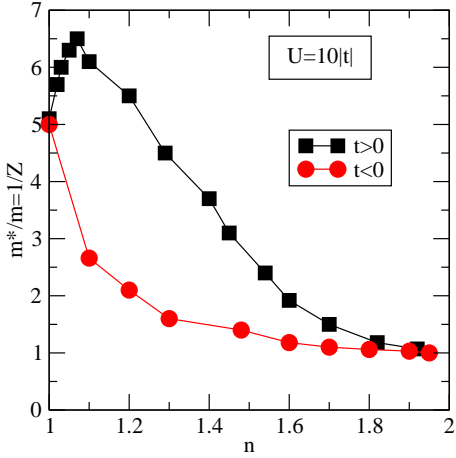}
\end{center}
\caption{(Color online.) The variation of the effective mass,
$m^*/m_b=1/Z$ with doping $n$ for both signs of $t$ for $U=10|t|$.
Notice in particular the  non-monotonic variation of the effective
mass with $n$ for $t>0$.} \label{fig10}
\end{figure}

In Fig. \ref{fig10} we show the variation
of the effective mass
with $n$ for $U=10|t|$. Strikingly the effective mass {\it
increases} when the system is doped off half-filling close to $n=1$
and $t>0$ displaying a maximum at about $n=1.07$. This is in
contrast to the $t<0$ case, and what has been found previously
for non-frustrated lattices such as the hypercubic lattice,
for which the effective mass decreases as the system
is doped away from half-filling, reflecting
reduced correlations.
 In fact for $n>1.07$ the effective masses do decrease with
increasing $n$, reaching the non-interacting limit ($Z \rightarrow
1$), as $n \rightarrow 2$. A non-monotonic behavior of the effective
mass is then found close to half-filling and $t>0$. This is in
contrast with the simple  behavior encountered for the hole doped
($n<1$) Mott insulator with a flat non-interacting DOS. In that
case, the effective mass behaves as\cite{Georges}
\begin{equation}
m^*/m_b \propto 1/x,
\end{equation}
diverging at the Mott transition. This result can also be derived
from slave-boson theory.\cite{fresard}

The effective mass increase with initial doping is a result of the
initial increase of the DOS for $t>0$ as $\epsilon_F$ moves closer
to the van-Hove singularity. The subsequent decrease in the
effective mass can be attributed to the decrease in the effective
number of charge carriers which suppresses correlation effects. It
would be interesting to see whether slave boson theory, with the DOS
for the triangular lattice, could reproduce the non-monotonic
dependence of the quasi-particle weight on doping. (This would
involve solving equations (8) and (9) in Ref. \onlinecite{fresard}).

\subsection{Sommerfeld-Wilson ratio}

\begin{figure}
\begin{center}
\epsfig{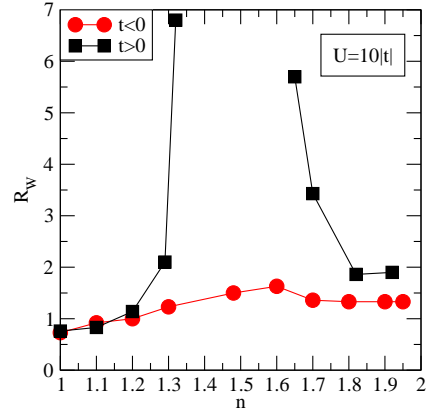}
\end{center}
\caption{(Color online.) The variation of the Sommerfeld-Wilson
ratio, $R_W=lim_{T\rightarrow0}(\chi(T)/\chi_0)/(\gamma(T)
/\gamma_0)$ with doping $n$ for both signs of $t$ for $U=10|t|$ and
$T = |t|/200$. } \label{fig11}
\end{figure}

In order to explore the effect of magnetic exchange in the metallic
correlated state  we analyze the Sommerfeld-Wilson
ratio,
\begin{equation}
R_W=\lim_{T \rightarrow 0}
{\chi(T)/\chi_0\over\gamma(T)/\gamma_0}={1 \over 1+F_0^a},
\end{equation}
where $F_0^a$ is the Fermi liquid parameter which is
a measure the
proximity of the system to a ferromagnetic
instability ($F_0^a=-1$ at the
instability) and $\gamma/\gamma_0=1/Z$.
The dependence of $R_W$ on the electron occupation factors are
plotted in Fig. \ref{fig11} for both $t>0$ and $t<0$.
 The ratio $R_W$ is the same at $n=1$ for
$t>0$ and $t<0$ but they behave very differently as the occupation is
increased. For the $t>0$ triangular lattice $R_W$ is strongly
enhanced signalling the proximity to a magnetic instability at about
$n=1.35$ and $n=1.65$ (see the phase diagram in Fig. \ref{fig1} for
$t>0$). For $t<0$, $R_W$ displays a maximum close to
$n=1.5$.

\subsection{Spectral density}

The different renormalization of the electrons for different signs
of $t$ is translated onto a different redistribution of spectral
weight induced by the Coulomb interaction as the spectral weight
suppressed at the Fermi energy must be transferred to higher
energies. We investigate this by computing local spectral
densities, $A_{\sigma}(\omega)=-{1 \over \pi} \text{Im}
G_{\sigma}(\omega+i \eta)$, where $\eta$ is an arbitrarily small
real number and
\begin{equation}
G_{\sigma}(i\omega)=\int_{-\infty}^{\infty} {d \epsilon
\rho(\epsilon) \over i \omega + \mu - \epsilon -\Sigma_{\sigma}(i
\omega)}.
\end{equation}

\begin{figure}
\begin{center}
\epsfig{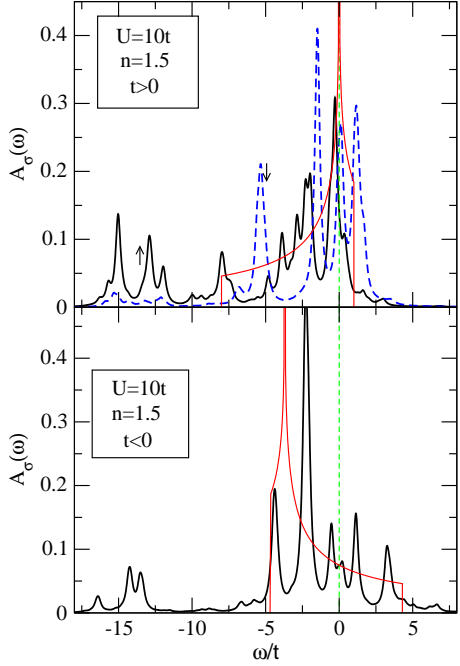}
\end{center}
\caption{(Color online.) Spectral densities for $U=10 |t|$ and
$n=1.5$ for the triangular lattice with $t<0$ and $t>0$. The
non-interacting DOS is also shown (red line) for comparison and
energies are relative to the Fermi energy. The majority ($\uparrow$)
and minority ($\downarrow$) spectral densities for $n=1.5$ and $t>0$
are shown as the system is ferromagnetic. } \label{fig12}
\end{figure}

In Fig. \ref{fig12} a comparison of the spectral
density $A_{\sigma}(\omega)$ with
the non-interacting DOS per spin, $\rho(\epsilon)$, is made for the case $n=1.5$ and $U=10|t|$. For
these parameters ferromagnetism occurs when $t>0$. Notice that
within the ferromagnetic metallic phase the spectral function of
majority spin subband ($\uparrow$) is found to be renormalized more
strongly than that of the minority spin subband ($\downarrow$). The
lower Hubbard band for spin up electrons contains more spectral
weight than the lower Hubbard band for spin down electrons. One may
think of this in terms of a photoemission experiment, clearly there
are many more spin up than spin down electrons  which can be
extracted from a ferromagnet. Alternatively, one may view this
effect as originating from the preclusion of the hopping of a
minority spin hole to a neighboring site by the presence of a
minority spin hole.

\subsection{Resistivity and Kadowaki-Woods ratio}

Within the DMFT approximation the conductivity is given
by,\cite{Georges}
\begin{equation}
\sigma(T)={e^2 \pi \over \hbar V} \int_{-\infty}^{\infty} d \omega
\left({-\partial f(\omega) \over \partial \omega}\right) \sum_{\bf k
\sigma} \left( {\partial \epsilon_{\bf k} \over \partial k_x }
\right)^2 A_\sigma({\bf k}, \omega)^2,
\end{equation}
where $V$ is the volume. In the limit of low temperatures, $T
\rightarrow 0$, the resistivity becomes $\rho \approx A T^2$, with
the A coefficient given by
\begin{equation}
A={\hbar \over e^2} {|C_{\Sigma}| \over \Phi(\tilde{\epsilon_F}) I}.
\end{equation}
The dimensionless constant $I=\int_{-\infty}^{\infty} {e^x \over
(x^2+\pi^2)(1+e^x)^2} \approx 1/12$, $C_{\Sigma}$ is the $\omega^2$
coefficient of the imaginary part of the self-energy, Im
$\Sigma(\omega) \approx -|C_{\Sigma}|(\omega^2+(\pi T)^2)$, and
$\Phi(\tilde{\epsilon_F})$ is given by
\begin{equation}
\Phi(\tilde{\epsilon_F})={1 \over V} \sum_{\bf k} \left({\partial
\epsilon_{\bf k} \over \partial k_x } \right)^2
\delta(\tilde{\epsilon_F}-\epsilon_{\bf k}).
\end{equation}
In the above equations
$\tilde{\epsilon_F}=\mu- {\rm Re}\Sigma(0)=\epsilon_F$, where $\epsilon_F$
is the Fermi energy of the non-interacting system\cite{Muller} for a
given electron occupation $n$. From the $T^2$ coefficient of the
resistivity, $A$, and the specific heat slope at low temperatures,
$\gamma=2 \pi^2 k_B^2 \tilde{A}(\tilde{\epsilon_F})/3=\pi^2 k_B^2
\rho(\epsilon_F)/Z$, where $\tilde{A}(\tilde{\epsilon_F})$ and
$\rho(\epsilon_F)$ are the interacting and non-interacting
quasiparticle density of states per spin. We can now obtain the
Kadowaki-Woods ratio,
\begin{equation}
{A \over \gamma^2} = { \hbar \over e^2} {9 |C_{\Sigma}| Z^2  \over
4 \Phi(\epsilon_F) \pi^4 k_B^2 I \rho(\epsilon_F)^2  }.
\label{eqn:KadoWoods}
\end{equation}

\begin{figure}
\begin{center}
\epsfig{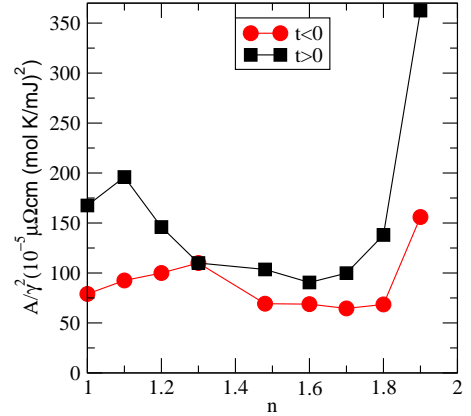}
\end{center}
\caption{(Color online.) The doping dependence of the
Kadowaki-Woods ratio $A/\gamma^2$ for $t<0$ and $t>0$ for fixed
$U=10|t|$. The lattice parameter, $a=2.84\AA$ relevant to Na$_x$CoO$_2$
has been used. Note the non-monotonic doping dependence of the ratio and
the fact that even at half-filling the ratios are different for the
different signs of $t$. } \label{fig13}
\end{figure}

The doping dependence of the Kadowaki-Woods ratio is plotted in
Fig. \ref{fig13} comparing the $t>0$ and $t<0$ cases. The figure
displays the non-monotonic dependence of the ratio with electron
occupation and also how different band structures can lead to
different absolute values of the Kadowaki-Woods ratio.

\section{Ferromagnetism}
\label{secferro}

Nagaoka rigorously proved \cite{Nagaoka} that the Hubbard model on a
connected lattice \cite{Tian} in the $U \rightarrow \infty$ limit
displays ferromagnetism when one hole (electron) is added to the
half-filled system when $t<0$ ($t>0$). This is due to the fact that
in this case, the kinetic energy is minimized if all the spins are
aligned in the same direction. This rigorous treatment has not been
extended to doping by more than one hole and it remains an
outstanding problem to further understand this interesting
phenomenon.\cite{Kollar}

\begin{figure}
\begin{center}
\epsfig{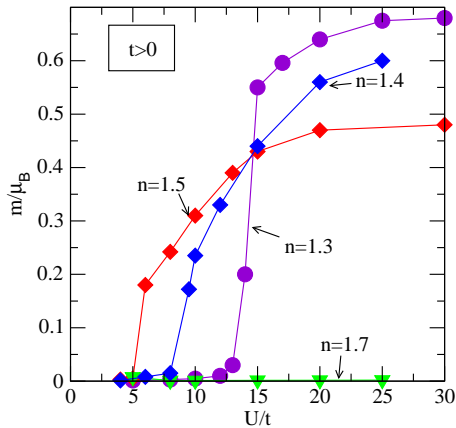}
\end{center}
\caption{(Color online.) Ferromagnetism in the electron doped $t>0$
triangular lattice from DMFT.  The magnitude of the spontaneous magnetic
moment is shown as a function of the electron occupation $n$.
As $U$ is increased  above $U_c(n)$ the system magnetizes spontaneously stabilizing
a ferromagnetic metal in a broad doping region. Ferromagnetism does not appear in
the electron doped $t<0$ triangular lattice.} \label{fig14}
\end{figure}

DMFT has proved to be an important tool to describe ferromagnetism
appearing due to local electronic correlations. The possibility of
Nagaoka or metallic ferromagnetism in a hole doped infinite
dimensional fcc lattice has been previously analyzed within
DMFT.\cite{Ulmke} More recently DFT(LDA)+DMFT calculations (where
DFT is density functional theory and LDA is the local density
approximation) have provided a realistic description of
ferromagnetism in Fe and Ni.\cite{Lichtenstein} Our calculations
show that for sufficiently large $U$ and $t>0$ ferromagnetism occurs
while the system is still metallic as the spectral density at the
Fermi energy is always finite for non-zero doping (c.f. Fig.
\ref{fig12}). In Fig. \ref{fig14} we show the sharp ferromagnetic
transition in the $T \rightarrow 0$ limit for $t>0$ obtained from
our DMFT calculations.

Increasing $U$ stabilizes the ferromagnetic region in a broader
electron occupation range. Ferromagnetism is found to be more stable
for  $1<n<1.5$ than for $1.5<n<2$, indicating the importance of
correlation effects as the system is closer to the Mott insulating
phase as $n \rightarrow 1$ and sufficiently large $U$. The
ferromagnetic transition is found to be sharp as shown in Fig.
\ref{fig14} even at the lowest $U$ value for $n=1.5$ where the van
Hove singularity occurs in the bare DOS. In contrast for $t<0$
where, for electron doping, there is no van Hove singularity we do
not observe ferromagnetism. This is, of course, rather reminiscent
of a Stoner type instability.

In order to explore the possibility of having Stoner ferromagnetism
we can obtain the critical value $U^S_c$ for ferromagnetism from the
Hartree-Fock (random phase approximation, RPA) Stoner condition
\begin{equation}
U^S_c(n) \rho(\epsilon_F)=1,
\end{equation}
where $\rho(\epsilon_F)$ is the bare DOS per spin. One would find
$U^S_c(n=1.5)=0$ and $U^S_c(n=1)=5$ for $t>0$ which are much smaller
values than the ones obtained from DMFT calculations (see Fig.
\ref{fig1}). Furthermore, from the Stoner condition we would also
expect a ferromagnetic instability at $U^S_c(n=1.3)=5.26$ and
$U^S_c(n=1.90)=10$ for $t<0$.  Clearly, simple Stoner ferromagnetism
is inconsistent with our DMFT results not only at the quantitative
but also qualitative level as the Stoner criterion would predict
ferromagnetism for both signs of $t$ for sufficiently large, but
finite, values of $U$. Stoner theory is known to overestimate
ferromagnetic tendencies.\cite{Vollhardt} In the present case, it
would predict ferromagnetism for $t<0$ in contrast to the more
sophisticated DMFT treatment. This difference can be attributed to
the on-site dynamical correlation effects which strongly
redistribute the spectral weight of the electrons and that are not
taken into account in standard mean-field theories.

Vollhardt {\it et al.}\cite{Vollhardt} have given a nice review of
the essential features of Hubbard models that are favourable towards
to ferromagnetism. (i) The energy dependence of the density of
states near the Fermi energy should be sufficiently asymmetric, with
the DOS being larger towards the top (bottom) of the band for
electron (hole) doping. A flat band system with a singular DOS at
the upper(lower) edge the most favorable for ferromagnetism. Then
the increase in electronic kinetic energy associated with spin
polarisation is smaller than for a constant DOS. (ii) Strong Coulomb
interactions narrow the bands. Less narrowing occurs for the
polarised case because polarisation reduces the effect of
correlations.

Our DOS is not of the flat band type, however,
the energy cost in completely polarizing the
non-interacting electron doped system for $t>0$ is found to be
much smaller than for $t<0$.
Indeed, for an electron doped system this is given by
\begin{equation}
\Delta E= 2 \int_{\mu_0}^{\epsilon_{max}} d \epsilon \epsilon
\rho(\epsilon) - \int^{\epsilon_{max}}_{\mu_p} d\epsilon \epsilon
\rho(\epsilon), \label{fullpol}
\end{equation}
which should be negative if ferromagnetism is stable. The chemical
potentials, $\mu_0$ and $\mu_p$, correspond to the unpolarized and
fully polarized systems, respectively and $\epsilon_{max}$ is the
energy at the upper edge of the band. Inserting $\rho(\epsilon)$ for
$t>0$ in expression (\ref{fullpol}) with $n=1.3$ we obtain $\Delta
E= 0.603 |t|$, to be compared with $\Delta E =1.34 |t|$ for $t<0$.
Interestingly, for the square lattice, $\Delta E= 0.8 |t|$, which is
between the values of $\Delta E$ for the $t<0$ and $t>0$ triangular
lattices. Hence, although unsurprisingly ferromagnetism is not
stable for the non-interacting system regardless of the sign of $t$,
the ferromagnetic tendencies are clearly stronger in the electron
doped triangular lattice for $t>0$ than they are for $t<0$ in
agreement with the conclusions derived from the three site cluster
analysis (sect. \ref{a ferromagnet} and Fig. \ref{fig2}).

Finally, it is worth comparing previous studies of the $t-J$ model
on the triangular lattice with our DMFT results. A Curie-Weiss metal
as well as ferromagnetism have been previously found from
calculations using high temperature series expansions, for the hole
doped triangular $t-J$ model.\cite{Ogata} Our results for the
electron doped lattice are equivalent to the hole doped case
discussed in Ref. \onlinecite{Ogata} once the sign of $t$ is
reversed. More specifically, ferromagnetism and Curie-Weiss metallic
behavior appear in the high temperature expansion for the hole doped
system with $t<0$ while a weak temperature dependent susceptibility
appears when $t>0$. Furthermore, a heavy fermion state, analogous to
our strongly correlated metallic state for $t>0$, is predicted to
exist in the hole doped system with $t<0$. The qualitative agreement
of the magnetic response obtained from DMFT compared with the high
temperature studies (which includes both local and non-local
correlation effects) of the $t-J$ model indicate the importance of
the local aspects of electronic correlations. This has also been
pointed out in the context of metallic magnetism for which a good a
description of the electronic and magnetic properties of Fe and Ni
is attained from a local theory such as DMFT. \cite{Lichtenstein}

\section{Comparison with experiments on $\textrm{Na}_x\textrm{CoO}_2$}\label{secexpt}

As this work has been largely motivated by experiments on \Na we now
consider what the lessons learned from a DMFT study of the Hubbard
model on a triangular lattice might have to say about \Nan. First,
our results show that the details of the electronic structure, and
in particular the DOS, play a crucial role in determining the
physics of strongly correlated frustrated systems such as \Nan. This
is something of a pyrrhic victory in that this very result tells us
that there is little hope of quantitative agreement between
experiments on \Na and calculations based on a simplified one band
models with nearest neighbour hopping only, such as those presented
above. Indeed our results do not show any such quantitative
agreement.

In particular for the triangular lattice with $t<0$ (which, of the two cases we
considered, gives the band structure closest to those
suggested by both ARPES and LDA-DFT) the DMFT of the electron doped Hubbard model
closely resembles a weakly interacting metal. This is clearly not
what is observed in \Nan, even at a qualitative level.

On the other hand many of the results for $t>0$ are qualitatively
consistent with the picture drawn by experiments on \Nan. In
particular we propose that the experimental `Curie-Weiss metal' is
little more than the ``bad metal'' in the extreme situation of very
strong electronic correlations and high frustration. Important
experimental support for this hypothesis comes from the fact that a
Fermi liquid like resistivity has only been observed\cite{Li} when
$T\lesssim1$~K which suggests that $T^*\sim1$~K. An obvious
objection, that $T^*$ predicted by DMFT will not be this small for
any reasonable parameters, will be discussed below (Sec.
\ref{spin-fluc}).

Our results for $t>0$ show that magnetism is a genuinely quantum
many-body effect, in particular the magnetism has little resemblance
to simple Stoner ferromagnetism. We propose that the observed A-type
antiferromagnetism results from in plane Nagaoka ferromagnetism with
a weak interlayer antiferromagnetic coupling. The
observation\cite{Bayrakci,Helme} that the strength of the effective
in-plane ferromagnetic coupling in \Na is the same order of
magnitude as the effective interlayer antiferromagnetic interaction
despite of the highly two-dimensional crystal- and band-structures
of \Na is naturally explained in this scenario as the effective
ferromagnetic interaction  in the Nagaoka is much weaker than the
intrinsic antiferromagnetic  interaction which gives rise to the
phase. This can be seen from Fig. \ref{fig10}, here $\theta(n)$
becomes positive and ferromagnetism is suppressed by reducing the
doping towards the critical doping which is the lower bound for
which ferromagnetism is observed. The intrinsic antiferromagnetic
interactions due to superexchange are given by $J=4t^2/U=0.4|t|$ for
$U=10|t|$, which we take in the calculations above. Therefore even
though $J=4t^2/U\gg J_\perp=4t_\perp^2/U$, where $J_\perp$ and
$t_\perp$ are respectively the inter-plane exchange constant and
hopping integral, near the phase transition to ferromagnetism
$J\gg\theta(n)$ and thus the inter-plane antiferromagnetic exchange
may be the same order of magnitude as the in-plane ferromagnetic
exchange, as is observed experimentally in \Nan.

To qualitatively test these ideas calculations with a more realistic
band structure are required. LDA calculations predict that the
dispersion of the bands is rather different to the dispersion
obtained from a nearest-neighbor isotropic triangular lattice. In
particular the LDA band is much flatter close to the $\Gamma$-point.
Indeed this is just the sort of change that is likely to have a
significant effect on the results of the type of calculations we
have presented above. In the current work we have also neglected the
possibility of Fermi surface pockets near the $K$-points arising
from the $e_{2g}^\prime$ band. \cite{Singh,LDA} Recall that, for
example, we demonstrated that the magnetism and Curie-Weiss metal
behavior we observed in the triangular lattice depend crucially on
the DOS \emph{near} to the Fermi level and not just on the DOS
\emph{at} the Fermi energy. Therefore, even if this band does not
actually cross the Fermi energy, as some studies have indicated,
\cite{Lee,ARPES} it may play an important role in \Nan.
Interestingly, the $e_{2g}^\prime$ band can be fitted to a $t>0$
tight-binding model close to the Fermi energy. Hence, the
$e_{2g}^\prime$ bands may play an important role in producing the
Curie-Weiss metallic state. Anyway, the generalization of these
results to multiband systems\cite{Ishida} is likely to be important
for a complete understanding on the magnetic and transport
properties of \Nan.

To quantify these remarks we now present a comparison of the DMFT
results for the triangular lattice with the experimentally measured
properties of \Nan.

\subsubsection{Electronic heat capacity}\label{secgamma}

The experimental \cite{Bruhwiler,Cv} value of the linear-$T$
coefficient of the specific heat ($C_v \approx \gamma T$ for $T<T^*$
with $T^*=1$ K) in Na$_{x}$CoO$_2$ and $0.7<x<0.82$ is
$\gamma=25-30$ mJ/(mol K$^2$) in zero magnetic field. This
corresponds to an effective mass enhancement $m^*/m_b \approx 3-4$
where $m_b$ is the LDA band mass. 
On applying an external magnetic field of $H=14$ T, $\gamma$ is
suppressed leading to $\gamma=0.02$ J/(mol K$^2$) (Ref.
\onlinecite{Bruhwiler}) implying at 20-30\% decrease in $m^*$.
The observed temperature and magnetic field
dependence also has two unusual features.
In zero field, $C_v(T)/T$ is non-monotonic, but
becomes monotonic in a field of 14 T
(see Fig. 3 in Ref. \onlinecite{Bruhwiler}).
The reduction of $C_v(T)/T$ as the temperature
increases from 1 to 5 K is consistent with the
destruction of quasi-particles in this temperature
range (compare Fig. 37 in Ref. \onlinecite{Georges})

Our DMFT calculations for $n=1.7$ and $U=10|t|$, predict a weak
renormalization of the electrons, $m^*/m_b \approx 1.1$, for $t<0$
and $m^*/m \approx 1.5$ for $t>0$. However, in the discussion to
follow it will be important to note that as the system is driven
towards half-filling, $n \rightarrow 1$, we find effective mass
enhancements which are comparable to or larger than
experimental values (see Fig. \ref{fig10}). 
Note that effective mass enhancement for $t<0$ is rapidly suppressed
(Fig. \ref{fig10}) as $n$ is increased above $n>1.2$ becoming
effectively a weakly correlated metal for this doping region. But,
for $t>0$ we find that the mass enhancement remains large to much
higher dopings than for $t<0$ (for $t>0$ the mass enhancement is
comparable or greater than that observed experimentally for
$x\lesssim 1.5$).

\subsubsection{Low-temperature resistivity}\label{secA}

Experimentally,\cite{Li} the resistivity varies as, $\rho(T) \approx
\rho_0 + A T^2$, as is expected for Fermi liquid, below $T\approx
1$~K. The coefficient of the quadratic term is measured\cite{Li} to
be $A=0.96~\mu\Omega$~cm/K$^2$ for Na$_{0.7}$CoO$_2$. On applying an
external magnetic field up to $H=16$ T,  the quadratic coefficient
of resistivity is decreased to $A=0.22~\mu\Omega$~cm/K$^2$ and $T^*$
is increased to 4 K.

DMFT predicts Fermi liquid behavior below $T^*$,  the low energy
coherence scale. The resistivity coefficient predicted by DMFT is
$A=0.0053~\mu\Omega$~cm/K$^2$ for $t<0$, $n=1.7$ and $U=10|t|$,
which is more than two orders of magnitude smaller than the
experimental value.  However, at half-filling and $U=10|t|$,
$A=0.81\mu \Omega$~cm/K$^2$, which is comparable to the experimental
result. For $t>0$, $n=1.7$ and $U=10|t|$ we find $A=0.32 \mu \Omega$
cm/K$^2$, which is the same order of magnitude as in experiments of
\Nan. Therefore, either the details of the DOS or driving the system
towards the Mott transition could be responsible for the large value
of $A$ observed experimentally. However it is important to stress
that in our calculations the coherence scale, $T^* \approx 100$ K
for $U=10|t|$ at half-filling, is two orders of magnitude larger
than the temperature below which $\rho(T) \approx \rho_0 + A T^2$ in
the experiments.\cite{Li} We will discuss the reasons for this below
(see Sec. \ref{spin-fluc}).

\subsubsection{Kadowaki-Woods ratio}

The Kadowaki-Woods ratio of Na$_{0.7}$CoO$_x$ is experimentally
found\cite{Li} to be $A/\gamma^2=60 a_0$, with $a_0=10^{-5} \mu
\Omega$ cm mol$^2$ K$^2$/mJ$^2$ being the constant value found in
the heavy-fermion materials. The large values of the ratio have been
recently discussed by Hussey,\cite{Hussey} who explained that when
volumetric rather than molar quantities for the specific heat, the
values of $A/\gamma^2$ are comparable to the ones in the heavy fermion
systems. From the DMFT calculations with $t<0$ we find that
$A/\gamma^2=64.2 a_0$ for $n=1.7$ whereas at half-filling
$A/\gamma^2=75 a_0$, implying a weak dependence with the electron
doping. A stronger dependence is found from DMFT for $t>0$
for which the ratios vary between $A/\gamma^2=100 a_0$
for $n=1.7$ and $170 a_0$ at half-filling (see Fig. \ref{fig13}).
Thus, the absolute values depend strongly on the DOS.

\subsubsection{Uniform magnetic susceptibility}

Experimentally, the magnetic susceptibility is found to have a
Curie-Weiss form, $\chi(T)\sim 1/(T+\theta)$, and a magnitude much
larger than the Pauli susceptibility expected for weakly interacting
metal across a large range of dopings that have metallic ground
states.\cite{Foo,Prabhakaran} Our DMFT calculations for $t<0$
display Pauli paramagnetism, which is typical of weakly correlated
metals, in qualitative disagreement with experiments. However, for
$t>0$ a Curie-Weiss metal is observed
across a wide range of dopings.
In the discussion that follows it will be important to recall that
the DMFT prediction of the susceptibility of a metal close to the Mott
transition is that it has the Curie-Weiss form with
$\theta=T^*+J>0$.
Thus, a behaviour similar to the experimentally observed Curie-Weiss metal is
predicted by DMFT for $t>0$ [i.e., when $\rho(\epsilon_F)$ is
large] and near to the Mott transition.


\subsubsection{Sommerfeld-Wilson ratio}

The experimentally measured Sommerfeld-Wilson ratio
\cite{Mukhamedshin} of Na$_{0.7}$CoO$_2$ is
$R_W=\lim_{T\rightarrow0}(\chi(T)/\chi_0)/(\gamma/\gamma_0)=7.8$.
For the hypercubic lattice, the Wilson ratio predicted by DMFT goes
to zero close to the Mott transition, i.e. at half-filling, but
becomes finite as soon as the system is doped away from
half-filling. \cite{Georges} On the triangular lattice with $t<0$ we
find that the Wilson ratio attains its largest values at $n=1.6$ at
which $R_W=1.7$ and decreases to 1.3 as $n \rightarrow 2$ 
with $U=10|t|$. In the limit $n \rightarrow 1$, the calculated $R_W
=0.8$. On the other hand for $t>0$, $R_W$ becomes extremely large
 at dopings close to the ferromagnetic instability.
For example, for $n \approx 1.65$, $R_W$ is already about 6. One
should note that experimentally, $x=0.7$ is extremely close to the
critical doping at which the A-type antiferromagnetism appears.
Therefore the large Sommerfeld--Wilson ratio observed is
probably a direct consequence of the proximity of Na$_{0.7}$CoO$_2$
to the magnetic transition. This analysis would also suggest that as
the doping is decreased from $x=0.7$ then $R_W$ will also decrease,
although calculations with a more realistic band structure are
required to confirm that DMFT makes this prediction.

\subsubsection{Effect of a small external
magnetic field on the low-T resistivity, $A$, and specific heat,
$\gamma$, coefficients}

A rapid suppression of the low-$T$ resistivity coefficient $A$ and
specific heat slope $\gamma$ and an increase in $T^*$ are observed
when an external small magnetic field is applied to \Nan. [The
details of the experimental observations of the behaviour of
$\gamma$ (Ref. \onlinecite{Bruhwiler}] and $A$ (Ref.
\onlinecite{Li}) are summarised in sections \ref{secgamma} and
\ref{secA} respectively.)

For the single impurity Kondo model
 the temperature dependence of the heat capacity has a maximum near
the Kondo temperature.
 In a magnetic field, once the Zeeman energy, $g \mu_B B$,
is comparable to the Kondo energy, $k_B T_K$, this maximum shifts to
higher temperatures.\cite{sacramento,Hewsonfig} Based on this we
would expect that the coherence temperature becomes larger with
increasing field.

DMFT for the Hubbard model in a magnetic field shows that in the
metallic phase of a weakly interacting metal the quasi-particle weight $Z$ increases with field,
consistent with the predictions of Stoner theory. (See Fig. 5 of
Ref. \onlinecite{Laloux}) Consequently, the specific heat coefficent
$\gamma$ will decrease with field. Similarily, the resistivity
coefficient $A$ will also decrease. In contrast, a strongly correlated metal
close to the Mott transition displays a weak suppression with magnetic field
(see inset of Fig. 9 in Ref. \onlinecite{Laloux}).


\subsubsection{Failure of one band models with nearest neighbour hopping}

The disagreements highlighted above between the one band model
with nearest neighbour hopping and experiments on \Nan, particularly
for $t<0$ which is the appropriate sign of $t$ for \Nan, does not
appear to be a result of the DFMT approximation. Our results are
entirely consistent with the series expansions\cite{Ogata} and
RVB\cite{Kumar} results for the $t-J$ model on a triangular lattice
in this regard. Therefore we believe that simple one
band models with nearest neighbour hopping only cannot account for
the observed behaviour of \Nan. This is an extremely important
result given the number of theoretical papers on both \Na and \NaH
that are based on this type of model. In the next section we examine
what kind of models or approximations are required to give an
account of the strong correlation effects in \Nan.

\section{More realistic models of
$\mathbf{Na_xCoO_2}$}\label{secbeyond}

\subsection{Beyond DMFT}

Although DMFT calculations on the triangular lattice suggest the
proximity of the metal to a Mott insulating phase several aspects of
the  observed low temperature behavior of \Na are still difficult to
understand. Foremost amongst these is why is the value of $T^*$
estimated from resistivity so much smaller than the obtained from
the susceptibility and DMFT?


The experimental value for the coherence temperature, $T^*=1$~K, as
determined from the temperature below which $\rho(T)=\rho_0+AT^2$,
is two orders of magnitude smaller than the DMFT result,
$T^*\approx0.1|t|\approx100$ K for $U=10|t|$ at half-filling. For
$t>0$ there is only a weak increase in $T^*$ with doping (see Fig.
\ref{fig5}). Driving the system closer to the Mott transition by
increasing $U$ would lead to a strong suppression of $T^*$, {\it
e.g.} $T^* \rightarrow 0$ as $U \rightarrow U_c$ at half-filling.
On the other hand, the values of $T^*$ predicted by DMFT are
comparable with values of $\theta \approx 150$ K found
experimentally\cite{Foo} from the Curie-Weiss fit of the spin
susceptibility, $\chi(T) \approx 1/(T+\theta)$, at large
temperatures.

Spin susceptibility experiments on Na$_{x}$CoO$_2$ show a clear
departure from the simple high temperature Curie-Weiss
susceptibility at temperatures of order the coherence temperature
predicted by DMFT (but orders of magnitude above the temperature
where the resistivity is observed to vary quadratically with
temperature). Below this temperature the spin susceptibility is
strongly enhanced in contrast to the DMFT prediction of a Pauli
susceptibility below $T^*$.
 Although
on this front one should beware of the major
differences\cite{CWM_foot} between the susceptibilities measured by
Foo \etal \cite{Foo} and those measured by Prabhakaran \etal
\cite{Prabhakaran}

\subsubsection{Spin fluctuations}\label{spin-fluc}

All of the results presented
in this paper were obtained from a purely local theory, DMFT.
Therefore these results neglect short range ferromagnetic
fluctuations which are likely to become increasingly important as
one dopes towards the ferromagnetic phase. Indeed such ferromagnetic
fluctuations have already been observed by neutron scattering in
Na$_{0.7}$CoO$_2$.\cite{Boothroyd,Helme,Bayrakci}

The importance of these fluctuations can be addressed within
Moriya's \cite{Moriya} spin fluctuation theory of the electronic and
magnetic properties of nearly ferromagnetic metals. The theory
assumes that the dynamic spin susceptibility has the form
\begin{eqnarray}
\chi({\bf q},\omega) = \frac {\chi({\bf q=0},\omega=0)}{1 +
\xi^2\left(q^2 - \frac{i \omega}{\Gamma q}\right) }
 \label{moriya1}
\end{eqnarray}
in two dimensions. $\xi$ is the ferromagnetic correlation length and
$\Gamma$ is a phenomenological parameter. This has the same
frequency and wavevector dependence as one obtains from an RPA
treatment of a Fermi liquid metal.

 Within this framework, the electronic
properties can be sensitive to an applied magnetic field as,
at low temperatures, it suppresses the ferromagnetic fluctuations,
driving the system into an unrenormalized Fermi liquid state.
 The low temperature scale $T_0$ is defined by\cite{Moriya}
\begin{equation}
T_0 = \frac{\Gamma q_B^3}{2 \pi}
 \label{moriya2}
\end{equation}
where $\pi q_B^2$ is a measure of the
area of the first Brillouin zone.

The spin susceptibility is enhanced as the temperature is lowered
departing from the Curie-Weiss law as a result of the enhancement
of ferromagnetic fluctuations at low temperatures. Also a small
magnetic field is expected to rapidly suppress $A$ and $\gamma$ as
experimentally observed in \Nan. The above arguments are
consistent with evidence of strong ferromagnetic fluctuations
acting in the Co-planes for $x\approx0.75$ from recent neutron
scattering experiments.\cite{Boothroyd} In order to determine the
importance of these spatial correlations it is desirable to
compare the measurements of $\xi(q,\omega)$ in Ref.
\onlinecite{Boothroyd} with the form (\ref{moriya1}). If at a
frequency $\omega$ one observes a peak of width $1/\xi$ in
wavevector space then $T_0 \sim \omega (q_B \xi)^3$. One sees from
Figures 1 and 4 in Ref. \onlinecite{Boothroyd} that for $\omega
\simeq 10$ meV that $q_B \xi \sim 0.4$. Hence, $ T_0 \sim 1-10$ K
and so the ferromagnetic fluctuations could be the origin of the
low energy scale in \Nan.

\subsection{Beyond the single band Hubbard
model}\label{sec:beyond-Hub}

\subsubsection{Role of ordering of the sodium ions}

We believe that the A-type antiferromagnetic phase observed
experimentally corresponds to the ferromagnetic phase predicted by
DMFT and many of the metallic state properties can be interpreted in
terms of their proximity to a ferromagnetic transition and strong
Coulomb repulsion effects. However, a number of experiments suggest
that \Na behaves like it is more strongly correlated as the doping,
$x$, is increased. These results are difficult to understand from a
simple model which only contains the Co planes as such models become
more weakly correlated as the electronic doping is increased due to
the small number of holes left in the Co-planes. Below we propose
that the ordering of the Na atoms observed at many dopings provides
a natural explanation of these effects as it can effectively drive
the system towards half filling by introducing a spatial modulation
of the site energy of the Co atoms. The suggestion that Na ordering
plays a crucial role that has not been appreciated until now is made
more plausible by the very sensitive dependence of the magnetic and
transport properties on the exact details of the DOS that we have
demonstrated above.

It seems likely that the charge ordering of the electrons in the
CoO$_2$ layers observed at $x=0.5$ is induced by Na
ordering.\cite{Foo} Recently, experimental studies by Zandbergen
\etal \cite{Na-expt} and Roger \etal \cite{Roger}, first principles
calculations by Zhang \etal \cite{Na_ordering} and classical
electrostatic calculations by Zhang \etal and Roger \etal have shown
that as $x$ is varied not only is the doping of the CoO$_2$ planes
varied, but so is the Na ordering. However, there remains debate
over the exact nature of the Na order at specific values of
$x$.\cite{Na-expt,Roger,Na_ordering} There is additional evidence
from NMR experiments of the presence of inequivalently charged Co
ions with Co$^{3+}$ (S=0) and Co$^{4+}$ (S=1/2) induced by the
charged Na layer of ions. \cite{Mukhamedshin} This effect implies
that a large fraction of the electrons in the Co planes behave as
$S=1/2$ localized moments. This is precisely the behavior expected
close to the Mott insulating phase in a Hubbard
model close to half-filling as described by DMFT. 

To make this proposal more concrete we analyse four special cases,
$x=$0.33, 0.5, 0.71 and 0.75. For each of these we take the
orderings proposed by Zhang \etal \cite{Na_ordering} and Zandbergen
\etal \cite{Na-expt} although we note that in two of these cases
($x=0.71$ and 0.75) Roger \etal have proposed different Na ordering
patterns which involve the formation of `vacancy droplets'. We
stress that this work has nothing to contribute to the debate over
which patterns are, in fact, realised and we take these patterns at
$x=0.71$ and 0.75 merely to exemplify our point. In the cases of
$x=0.33$ and 0.5 we find that the Na ordering introduces two Co
sublattices while for $x=0.71$ we find three Co sublattices and for
$x=0.75$ we find four Co sublattices. For this analysis one needs to
consider the two possible locations of Na in \Nan, referred to as
Na(1) and Na(2).\cite{Na-expt} The CoO$_2$ planes are at $z=0.25c$,
where $c$ is the lattice constant perpendicular to the cobaltate
layers. The Na(1) sites lie directly above (at $z=0.5c$) and below
($z=0$) the Co atoms, while the Na(2) sites are above or below the
centres of the triangles in the triangular lattice formed by the Co
atoms. The simplest model which can include the effects Na ordering
is a Hubbard model which includes the Co site energy modulated by
the number of nearest Na sites that are occupied. Clearly, one
expects that occupied Na(1) and Na(2) will modulate the Co site
energy by different amounts which we denote $\varepsilon_1$ and
$\varepsilon_2$ respectively. If $\mu$ is the chemical potential for
$x=0$ the Hamiltonian is then given by
\begin{eqnarray}
H &=& \sum_{i,\sigma}\;(N_{1i}\varepsilon_1+N_{2i}\varepsilon_2-\mu)n_{i\sigma}  \label{eqn:order}\\
&&-t\sum_{<ij>,\sigma}\;(c^\dag_{i\sigma} c_{j \sigma} + H.c.)
\notag + U \sum_{i} n_{i\uparrow} n_{i\downarrow}
\end{eqnarray}
where $N_{1i}$ and  $N_{2i}$ give the number of occupied nearest
neighbour Na(1) and Na(2) sites respectively. This model assumes
that the electronic degrees of freedom relax on much shorter time
scales than the Na ions move, which is reasonable because of the
relative masses, even though the effective quasiparticles mass is an
order of magnitude larger than the bare electronic mass.

\begin{figure}
\begin{center}
\epsfig{file=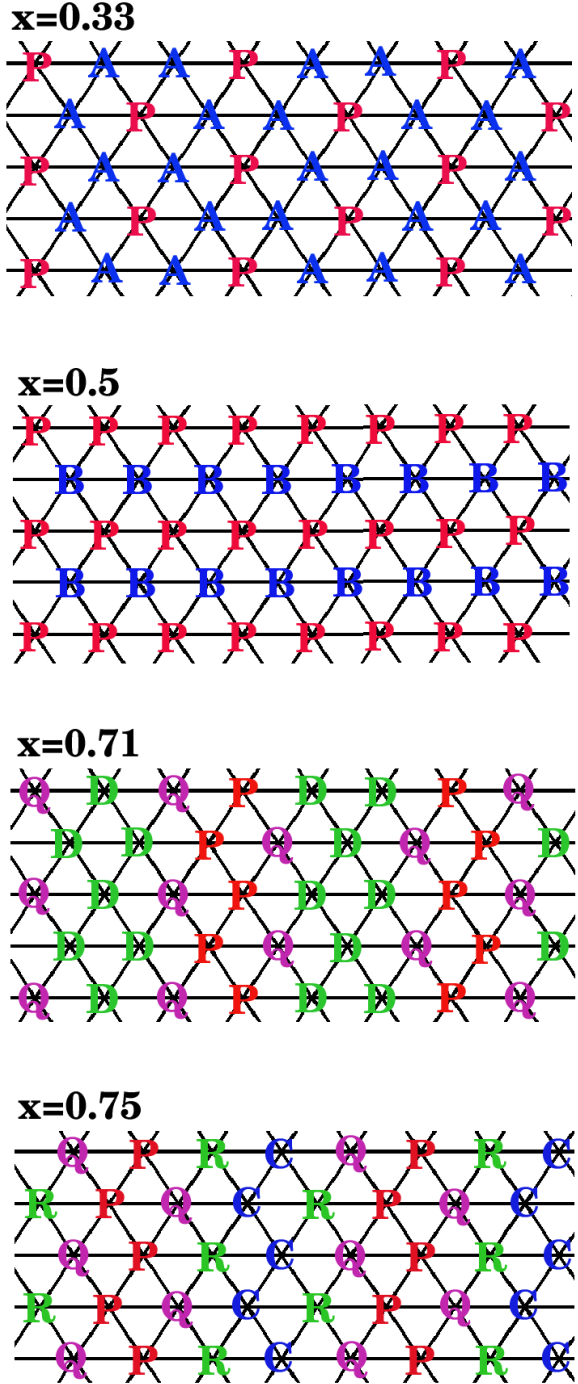,width=8.cm,angle=0,clip=}
\end{center}
\caption{(Color online.) Different sublattices of cobalt ions due to
commensurate ordering of sodium ions. Pictorial representation of
the Hamiltonians given by eqn. (\ref{eqn:order}) are shown for
$x=0.33$, 0.5, 0.71 and 0.75. The different letters denote the
different site energies defined by eqn. (\ref{eqn:site-energy}), so
that, for example a P represents a site with site energy
$\varepsilon_P$. These different site energies result from the
different Na order patterns, which have been observed
experimentally\cite{Na-expt} and predicted from first principles
calculations. \cite{Na_ordering} The Na ordering for the quoted
filling factors are taken to be those shown in Fig. 2 of Ref.
\onlinecite{Na_ordering}. Note that Roger \etal\cite{Roger} proposed
different ordering patterns for $x=0.71$ and 0.75, which involved
`vacancy clustering'.} \label{fig15}
\end{figure}

To simplify our notation we now introduce the following
nomenclature,
\begin{eqnarray}
\varepsilon_A &=& \varepsilon_2-\mu \notag\\
\varepsilon_B &=& 2\varepsilon_2-\mu \notag\\
\varepsilon_C &=& 3\varepsilon_2-\mu \notag\\
\varepsilon_D &=& 4\varepsilon_2-\mu \label{eqn:site-energy}\\
\varepsilon_P &=& \varepsilon_1+ \varepsilon_2-\mu\notag\\
\varepsilon_Q &=& \varepsilon_1+ 2\varepsilon_2-\mu\notag\\
\varepsilon_R &=& \varepsilon_1+ 3\varepsilon_2-\mu\notag
\end{eqnarray}
This allows us to sketch the relevant lattices in Fig. \ref{fig15}.
It can be seen that for $x=0.33$ we have a honeycomb lattice of
sites with site energy $\varepsilon_A$ with the central sites in the
lattice having site energy $\varepsilon_P$. For $x=0.5$ the lattice
contains alternating chains of sites with site energy
$\varepsilon_B$ and $\varepsilon_P$. With $x=0.71$ we find the
lattice contains sites with site energies $\varepsilon_D$,
$\varepsilon_P$ and $\varepsilon_Q$ with all three sublattices
forming stripes. For $x=0.75$ a complicated arrangement of
interlaced stripes of sites with site energies $\varepsilon_C$,
$\varepsilon_P$, $\varepsilon_Q$ and $\varepsilon_R$ is formed. This
model (\ref{eqn:order}) will clearly make very different predictions
from the Hubbard model if the variations in the site energies caused
by Na ordering are sufficiently large. For example for $x=0.5$ and
$\varepsilon_P\gg\varepsilon_B$ (as one might expect) and $U\sim W$
a charge ordered insulator is likely to be the ground state of
Hamiltonian (\ref{eqn:order}), this is the state observed in
Na$_{0.5}$CoO$_2$. Roger \etal have estimated from classical
electrostatics that $\Delta\varepsilon\equiv
\varepsilon_P-\varepsilon_B\simeq100$~meV which is consistent with
this scenario. Further this kind of effect will drive the system
closer to Mott insulating phase if the occupation of one or more of
the sublattices is strongly suppressed. Thus we expect that the DMFT
treatment of (\ref{eqn:order}) will have significantly better
qualitative agreement with the experimentally observed properties of
\Na than the DMFT treatment of the Hubbard model on a triangular
lattice does. Finally we note that, if one of the sublattices of
$x=0.5$ can be `integrated out' of the effective low energy theory
due to a large disparity in the site energies we are left with a
Mott insulator on a rectangular lattice. Along one side (the
horizontal direction in Fig. \ref{fig15}) of the rectangle it can
be seen that the antiferromagnetic exchange constant is $J=4t^2/U$.
In a similar manner superexchange leads to an exchange constant
\begin{eqnarray}
J'= \frac{16t^4}{\Delta\varepsilon^2}\left[\frac{1}{U} +
\frac{1}{2\Delta\varepsilon+U} + \frac{1}{2\Delta\varepsilon}
\right]
\end{eqnarray}
in the perpendicular direction. Additionally there is a (diagonal)
next nearest neighbour exchange interaction
\begin{eqnarray}
J''= \frac{4t^4}{\Delta\varepsilon^2}\left[\frac{1}{U} +
\frac{2}{2\Delta\varepsilon+U} \right].
\end{eqnarray}
It is interesting to consider the three limiting cases: 
%
\begin{eqnarray}
&&{\rm for}~U\gg\Delta\varepsilon,~
\frac{J'}{J}=\frac{2t^2U}{(\Delta\varepsilon)^3}~ {\rm and}~
\frac{J''}{J'}=\frac{3\Delta\varepsilon}{2U}\ll1; \notag
\\&& {\rm
for}~ U\ll\Delta\varepsilon,~
\frac{J'}{J}=\frac{4t^2}{(\Delta\varepsilon)^2}\ll1 ~{\rm and}~
\frac{J''}{J'}=4; \notag
\\&& {\rm for}~ U=\Delta\varepsilon,~
\frac{J'}{J}=\frac{6t^2}{U^2}\ll1 ~{\rm and}~
\frac{J''}{J'}=\frac{5}{18}. \notag
\end{eqnarray}
In particular note that for $U\gg\Delta\varepsilon$ we may have
$J'\sim J$. Neutron scattering experiments suggest that the charge
ordered phase of Na$_{0.5}$CoO$_2$ has long range N\'eel
order\cite{Yokoi} which suggests $J\sim J'>J''$ and is therefore
consistent with this model when $U\gg\Delta\varepsilon\gg t$, which
is precisely the regime that the separate estimates of these three
parameters suggest we are in.

Before moving on from the effects of Na ordering it is interesting
to note that this type of problem is actually rather general in
doped systems. For example, in most models of the cuprates it is
assumed that the only effect of varying the doping is to vary the
electronic density in the CoO$_2$ layers. However, if it is
confirmed that Na ordering plays the role we have proposed in \Na
then this may be an important step towards a general understanding
of the role of dopant impurities in strongly correlated systems.

We stress that the above proposal relies on Na ordering and not
disorder. It is well known that disorder only plays a significant
role in DMFT when there are rather extreme levels of
disorder.\cite{Vlad} Therefore one would not expect our results to
be significantly affected by the low levels of disorder typically
found in \Nan. However, recently the role of disorder on the band
structure of \Na has received some attention.
\cite{Singh-disorder} If disorder does induce significant changes in
the DOS this would be an important factor for the results reported
in this paper. More generally, the role of disorder in \NaH will no
doubt prove an important probe of whether the superconducting
state is conventional or not.\cite{disorder}

\subsubsection{Multiple bands}

The model, as written in eqn. (\ref{eqn:order}), only contains a
single band. As discussed in the above band structure calculations
suggest a simple one band model is not sufficient to describe the
band structure of \Nan. Although the a$_{1g}$ band can
be roughly fitted to a tight-binding model with $t<0$ which is
consistent with the large hole pocket around the $\Gamma$ point,
additional small hole pockets may be present in the Fermi surface
associated with the e$_{2g}^\prime$ bands crossing the Fermi energy.
These bands can be fitted to a $t>0$ tight-binding band which is
likely to favour ferromagnetism. Such a multi-band model is a
straightforward generalisation of (\ref{eqn:order}) giving
\begin{eqnarray}
H &=&
\sum_{i\sigma\nu}\;(N_{1i\nu}\varepsilon_1+N_{2i\nu}\varepsilon_2-\mu_\nu)n_{i\nu\sigma}
\notag \\   \label{eqn:order2}
&&-\sum_{ij\nu\nu'\sigma}\;t_{ij\nu\nu'}(c^\dag_{i\nu\sigma}
c_{j\nu'\sigma} + H.c.) \\ \notag && + \sum_{i\nu\nu'}U_{\nu\nu'}
n_{i\nu\sigma} n_{i\nu'\sigma'}
\end{eqnarray}
where the labels $\nu$ and $\nu'$ refer the different orbitals. A
multiband DMFT treatment is likely to describe both of the following
effects: (i) the proximity of the metal to a Mott insulating phase
leading to renormalized quasiparticles and Curie-Weiss behavior for
$T>T^*$ and (ii) the strong ferromagnetic fluctuations appearing at
very low temperatures. Several models have been recently proposed to
describe \Nan. In comparison to single band \cite{Kumar,Motrunich}
and multiband Hubbard models \cite{Ishida,JohannesEPL,LDA}, our
proposed model explicitly contains the periodic potential of the Na
layers acting on the Co planes in combination with the Coulomb
interaction. This is an important ingredient to understand the
unexpected Curie-Weiss susceptibility and large effective mass at $x
\rightarrow 0.7$. At these large dopings the system is expected to
behave as a weakly correlated metal.  We believe that our model is
sufficient to fully understand the interplay between these two
essential ingredients of the problem.

The problem of applying DMFT to a Hubbard model with two
inequivalent sublattices has been considered before. Chitra and
Kotliar\cite{chitra} considered the Hubbard model at half filling in
the presence of antiferromagnetic order. This leads to a different
chemical potential on the two sublattices. Bulla and
collaborators\cite{bulla}  considered charge ordering in the
extended Hubbard model at one-quarter filling. For the case of the
Bethe lattice the model was mapped to a pair of Anderson impurity
models, one for each of the two sublattices. For $V$ (the Coulomb
repulsion between electrons on neighbouring lattice sites) larger
than about $t$ and $U=2t$ the quasi-particle weight (and presumably
the coherence temperature also) is orders of magnitude smaller than
the non-interacting value and there is a pseudogap in the density of
states. Based on the above we would expect that for $x=0.5$ the
ground state will be insulator if $|\varepsilon_P - \varepsilon_B|
\sim U > |t|$ due to the two different sublattices.

\section{Conclusions}
\label{seccon}

We have presented a DMFT analysis of the electronic and magnetic
properties of the doped isotropic triangular Hubbard model. An important
result  of our work is the large effect of the bare DOS on its
magnetic and electronic properties. In particular, a
Curie-Weiss metal is found for $t>0$ and $U \gtrsim W$. In contrast,
Pauli paramagnetism is found for $t<0$ regardless of the magnitude
of $U$. We find a larger degree of localization and a larger
renormalization of the electrons for $t>0$ than for $t<0$ due to
the different DOS at the Fermi energy. The  spin susceptibility
crosses over from Curie-Weiss behavior, at large temperatures,  to a
renormalized Fermi liquid as $T \rightarrow 0$. At low temperatures,
the uniform spin susceptibility  is strongly enhanced for $t>0$  due
to the proximity to the ferromagnetic transition. This is in
contrast to the behavior for $t<0$, for which only a small
enhancement of $\chi(T \rightarrow 0)/\chi_0$ is found. The stronger
renormalization of the quasiparticles found for $t>0$ than for $t<0$
when the system is doped away from half-filling is a consequence of
a larger DOS near the Fermi energy in the $t>0$ case.

For $t>0$ we find a metallic ferromagnetic state when $U$ is
sufficiently large. The ferromagnetism is significantly different
from Stoner ferromagnetism in that it has a strong local moment
character and the criterion for ferromagnetism is not simply related
to the DOS at the Fermi level, but depends on the features of the
DOS over a significant energy range around the Fermi level. No
ferromagnetism has been found for any band filling or any value of
$U$ when $t<0$, which is qualitatively different from the prediction
of Stoner theory. We conclude that the ferromagnetism we have
observed is much more closely related to Nagaoka ferromagnetism. The
different behaviour of the two lattices is due to kinetic energy
frustration which dramatically reduces the energy cost of polarizing
the system for $t>0$. We have therefore proposed that the A-type
antiferromagnetic phase observed in \Na for $x>0.75$ is the result
of in plane Nagaoka type ferromagnetism.

The different behaviors encountered away from half-filling for the
different signs of $t$ contrasts with the system sufficiently
close to the Mott insulating phase ($n \rightarrow 1$ and $U>U_c$ or
$U \rightarrow  U_c$ and  $n=1$). In this case, the model displays
very similar properties including Curie-Weiss susceptibility with
$\theta>0$ (due to antiferromagnetic superexchange interaction close
to half-filling), a large effective mass enhancement, ($m^*/m_b=5$
for $U=10|t|$), large low-$T$ resistivity coefficient, and enhanced
Kadowaki-Woods ratio.

Thus our results show that the details of the band structure are
extremely important for understanding the metallic and spin ordered
phases of \Nan. In particular, a simple one-band Hubbard model on a triangular lattice
is not enough to understand the unconventional properties of \Nan.
Our results however do suggest that the Curie-Weiss susceptibility observed in
the metallic phase of \Na results from incoherent quasi-localized
electrons. Remarkably, the DMFT estimate of $T^*=0.1|t|\approx
100-200$~K is in good agreement with the Curie-Weiss constant
$\theta \approx 150$~K experimentally measured.

Many of the thermodynamic and transport properties
observed in \Nan, for large $x$, are consistent with the DMFT
prediction of the transport and magnetic properties of the metal
very close to the Mott transition. We have proposed that Mott
physics is relevant to Na$_{x}$CoO$_2$ even when $x$ is large
and the system is far away from the Mott insulating phase 
because of experimental observation of charge ordering in
the Co planes.

The $T$ dependence of several transport and magnetic properties is
consistent with the DMFT description of a metal close to the Mott
transition for $T>T^*$. However, at very low temperatures,
experimental observations on Na$_{x}$CoO$_2$, depart from our DMFT
results. For instance, the susceptibility deviates significantly
from the Curie-Weiss behavior being strongly enhanced below $T \sim
80$ K. Also transport and thermodynamic quantities are very
sensitive to a small applied magnetic field. One is tempted to
associate the very low temperature scale of about $1$~K
experimentally determined to the DMFT coherence scale, $T^*$, which
goes to 0 as the system approaches the Mott insulating state.
However, such a small value of $T^*$ implies an extremely large
effective mass enhancement in clear disagreement with experimental
values, $m^*/m_b \sim 3-4$. We conclude that this very small energy
scale is due to low energy ferromagnetic fluctuations in the
Co-planes which are neglected within the DMFT. These fluctuations
would scatter the strongly renormalized quasiparticles which are
properly described by DMFT.

We have then used this analysis to help identify the simplest
relevant model that captures the essential physics in \Nan. We have
proposed a model which contains the charge ordering phenomena
observed in the system that we have proposed drive the system close
to the Mott insulating phase for the large dopings. Band structure
calculations find that the a$_{1g}$ band can be roughly fitted to a
tight-binding model with $t<0$ which roughly describes the large
hole pocket around the $\Gamma$ point. Additional small hole pockets
may be present in the Fermi surface associated with the e$_{2g}^\prime$
bands crossing the Fermi energy. These bands can be fitted to a
$t>0$ tight-binding band for which ferromagnetic fluctuations are
likely to exist. Hence, a multiband DMFT treatment containing the
a$_{1g}$ and e$_{2g}^\prime$ bands as well as charge ordering phenomena
should describe both the following aspects: (i) the proximity of the
metal to a Mott insulating phase leading to renormalized
quasiparticles and Curie-Weiss behavior for $T>T^*$ and (ii) the
strong ferromagnetic fluctuations appearing at low temperatures.
We believe that a model including both of these ingredients is
needed to fully understand the physics of \Nan.

\acknowledgments

It is a pleasure to acknowledge helpful discussions with A.
Boothroyd, R. Coldea, J.O. Fj\ae restad, C. Hooley, M.D. Johannes,
A. Levy-Yeyati, B. Lake, M. Long, S. Nagler, F. Rivadulla, A.J.
Schofield, D.J. Singh and D.A. Tennant. We thank the Clarendon
Laboratory, Oxford University and the Rutherford Appleton Laboratory
for hospitality. J. M. is grateful to the Ram\'on y Cajal program
from Ministerio de Ciencia y Tecnolog\'ia in Spain and EU through
contract MERG-CT-2004-506177 for financial support. B.J.P. and
R.H.M. were supported by the Australian Research Council. Some of
the calculations where performed at the Scientific Computational
Center of Universidad Aut\'onoma de Madrid.


\begin{thebibliography}{}
\bibitem{ET-exp}
BEDT-TTF is bis(ethylenedithio)-tetrathiafulvalene.
\bibitem{foot-def}
To completely define our states we must also specify the order in
which they are filled. In what follows we fill the states in
numerical order of site number with $\uparrow$ states for a given
site filled before $\downarrow$ states. Thus we have, for example,
$|\uparrow\downarrow,\uparrow\downarrow,0\rangle =
c_{2\downarrow}^\dagger c_{2\uparrow}^\dagger
c_{1\downarrow}^\dagger c_{1\uparrow}^\dagger|0\rangle$ and
$|\uparrow\downarrow,\uparrow,\downarrow \rangle =
c_{3\downarrow}^\dagger c_{2\uparrow}^\dagger
c_{1\downarrow}^\dagger c_{1\uparrow}^\dagger |0\rangle$. This
ensures that the singlets are negative superpositions and the
triplets are positive superpositions which is intuitive. Note that
this is not the case for other filling conventions, e.g. if we fill
the $\uparrow$ states in numerical order and then the $\downarrow$
states in numerical order the state $c_{3\downarrow}^\dagger
c_{1\downarrow}^\dagger c_{2\uparrow}^\dagger c_{1\uparrow}^\dagger
|0\rangle + c_{2\downarrow}^\dagger c_{1\downarrow}^\dagger
c_{3\uparrow}^\dagger c_{1\uparrow}^\dagger |0\rangle$ is a singlet.
\bibitem{foot-silly}
The $U\rightarrow0$ limit of eqn (\ref{eqn:E-}) is slightly
confusing as one must remember that $t<0$. Using Eq. (\ref{eqn:E-})
we get $E(U\rightarrow0) =(1/2)\times(-2|t| - \sqrt{36|t|^2})-4|t|$ as required.
\end{thebibliography}
\end{document}